\documentclass[aps,pra,twocolumn,superscriptaddress,groupedaddress]{revtex4}

\usepackage{amssymb} \usepackage{color,graphicx} \usepackage{amsmath}
\usepackage{amsbsy} \usepackage{amsthm} \usepackage{bbm}
\usepackage{bm,bbm} \usepackage{float} \usepackage{braket}
\usepackage{placeins}
\usepackage[colorlinks=true,citecolor=blue,linkcolor=red,urlcolor=red]{hyperref}
\usepackage[sort&compress]{natbib}
\usepackage{comment}
\usepackage{stackengine,rotating,multirow}

\usepackage{array}
\newcolumntype{C}{>{$\displaystyle} c <{$}}

\newcommand{\bq}{\begin{equation}} \newcommand{\eq}{\end{equation}}
\newcommand{\bqali}{\bq\begin{aligned}}
\newcommand{\eqali}{\end{aligned}\eq}
\newcommand{\bqn}{\begin{equation*}}
\newcommand{\eqn}{\end{equation*}}

\newcommand\Dd[1]{\operatorname{d}^{#1}\!}
\newcommand\D{\operatorname{d}\!}
\renewcommand\k{{\bf k}}

\newcommand\p{{\bf p}}
\newcommand\z{{\bf z}}

\newcommand\q{\bf q}
\newcommand\x{\bf x}

\newcommand\F{{\bf F}}
\newcommand\kb{k_\text{\tiny B}}

\newcommand\erf{\operatorname{erf}}
\newcommand\com[2]{[#1,#2]}
\newcommand\acom[2]{\{#1,#2\}}

\newcommand\gammam{\gamma_\text{\tiny m}}
\newcommand\Scsl{\mathcal S_{\text{\tiny CSL}}}
\newcommand\DNS{\mathcal S_z(\omega)}
\newcommand\besseli{\operatorname{I}}
\newcommand\lcsl{\lambda}
\newcommand\rC{r_\text{\tiny C}}
\newcommand\muA{\mu_\text{\tiny A}}
\newcommand\muB{\mu_\text{\tiny B}}
\newcommand\rhos{\hat \rho}

\newcommand\Nlay{{N_\text{\tiny lay}}}

\graphicspath{{../Figures/}}

\begin{document}

\author{Matteo Carlesso}
\email{matteo.carlesso@ts.infn.it}
\affiliation{Department of Physics, University of Trieste, Strada Costiera 11, 34151 Trieste, Italy}
\affiliation{Istituto Nazionale di Fisica Nucleare, Trieste Section, Via Valerio 2, 34127 Trieste, Italy}

\author{Andrea Vinante}
\affiliation{Department of Physics and Astronomy, University of Southampton, SO17 1BJ, UK}
\affiliation{Istituto di Fotonica e Nanotecnologie, CNR - Fondazione Bruno Kessler, I-38123 Povo, Trento, Italy}
\author{Angelo Bassi}
\affiliation{Department of Physics, University of Trieste, Strada Costiera 11, 34151 Trieste, Italy}
\affiliation{Istituto Nazionale di Fisica Nucleare, Trieste Section, Via Valerio 2, 34127 Trieste, Italy}

\title{Multilayer test masses to enhance the collapse noise}

\date{\today}
\begin{abstract}

Recently, a non-thermal excess noise, compatible with the theoretical prediction provided by collapse models, was measured in a millikelvin nanomechanical cantilever experiment [Vinante \emph{et al}., Phys.~Rev.~Lett.~\textbf{119}, 110401 (2017)]. We propose a feasible implementation of the cantilever experiment able to probe such a noise. The proposed modification, completely within the grasp of current technology and readily implementable also in other type of mechanical non-interferometric experiments, consists in substituting the homogeneous test mass with one composed of different layers of different materials. This will enhance the action of a possible collapse noise above that given by standard noise sources.

\end{abstract}
 \maketitle

\section{Introduction}

Technological development allows for novel and more refined tests of the foundations of quantum mechanics \cite{Sinha:2010aa,Hensen:2015aa,Procopio:2017aa}, which were wishful thinking up to a few decades ago. Among them, non-interferometric tests \cite{Kovachy:2015ab,Usenko:2011aa,Vinante:2006aa,Abbott:2016ab,Abbott:2016aa,Armano:2016aa,Armano:2018aa,Aalseth:1999aa,Adler:2018aa,Bahrami:2018aa,Vinante:2016aa,Vinante:2017aa,Carlesso:2016aa,Helou:2017aa,Piscicchia:2017aa} of models of spontaneous wave function collapse \cite{Bassi:2003aa, Bassi:2013aa}, which assume a progressive violation of the quantum superposition principle when moving from the micro to the macro scale, have given a strong boost to the search of the limits of validity of quantum theory. These limits, if present, would represent an intrinsic boundary to the scalability of quantum technologies.

Collapse models predict the existence of new effects, which tend to localize the wave function of massive systems in space. This is accomplished by coupling quantum systems nonlinearly to a noise field, which is characterized by two phenomenological constants: a collapse rate $\lcsl$ and a correlation length $\rC$. Numerical value for these parameters were first given by Ghirardi, Rimini and Weber (GRW)~\cite{Ghirardi:1986aa}: $\lambda=10^{-16}$\,s$^{-1}$ and $\rC=10^{-7}$\,m. Later, Adler~\cite{Adler:2007ab,Adler:2007ac} suggested stronger values for the collapse rate, namely  $\lambda=10^{-8\pm2}$\,s$^{-1}$ for $\rC=10^{-7}$\,m, and $\lambda=10^{-6\pm2}$\,s$^{-1}$ for $\rC=10^{-6}$\,m. 

The literature on experimental tests of collapse models is nowadays rather extensive. First came matter-wave interferometry---the most natural type of experiment---where larger and larger systems are prepared in delocalized states, and quantum interference is measured by standard interferometric techniques~\cite{Eibenberger:2013aa,Hornberger:2004aa,Toros:2017aa,Toros:2018aa,Lee:2011aa,Belli:2016aa}. Due to the difficulty in handling massive delocalized states, such experiments so far do not place significant bounds on the collapse parameters. 

To overcome this difficulty, non-interferometric experiments have been developed. They are based on an unavoidable side-effect of the collapse process:  a diffusion of the system's position, which can be traced via optomechanical techniques, being these very sensitive to small position displacements~\cite{Bahrami:2014aa,Nimmrichter:2014aa,Diosi:2015ab}. Among them, cold atoms \cite{Bilardello:2016aa}, measurement of bulk temperature \cite{Adler:2018aa,Bahrami:2018aa} and detection of spontaneous X-ray emission give the strongest bound on $\lcsl$ for $\rC<10^{-6}$\,m \cite{Piscicchia:2017aa}, while force noise measurements on nanomechanical cantilevers \cite{Vinante:2016aa,Vinante:2017aa} and on gravitational wave detectors give the strongest bound for $\rC>10^{-6}$\,m \cite{Carlesso:2016aa,Helou:2017aa}. {Recently an excess noise of unknown origin was measured in one such experiment~\cite{Vinante:2017aa}, and several standard explanations were ruled out. The result is still unconfirmed and could be likely explained by more subtle conventional effects. Nevertheless one cannot rule out non-standard explanations, such as collapse models or decoherence effects due to the interaction with exotic particles or forces \cite{Bassi:2017aa}. In particular the fact that the noise is compatible with CSL collapse rate predicted by Adler \cite{Adler:2007ab,Adler:2007ac} calls for more sensitive experimental tests of collapse models.}

We propose a method to enhance and optimize the CSL effect in optomechanical setups, which can be readily applied to most experiments of this kind. In contrast with other previous proposals \cite{Goldwater:2016aa,McMillen:2017aa,Schrinski:2017aa,carlesso:2018ab}, the hereby described method takes the advantage of only existing technology, that was already used to set bounds on the CSL parameters. It consists in using a mechanical test mass composed of layers of two different materials, instead of an homogeneous one. {A similar technique was already considered for coherently enhancing weak quantum effects, see for example \cite{Smith:1984aa}.}
 For specific values of ratio between the layers thickness and $\rC$, the CSL noise coherently correlates the collapses of the single layers and an amplification mechanism, which is fully discussed below, emerges.
We will consider a specific application to the cantilever-based experiment described in Ref.~\cite{Vinante:2017aa}. The foreseen increase of the CSL effect is sufficient to test almost the entire interval of collapse rate proposed by Adler, and in particular to falsify the hypothesis that the excess noise observed in \cite{Vinante:2017aa} may be due to CSL. \\

\section{Model}

The CSL master equation~\cite{Bassi:2003aa}  is of the Lindblad type: $\D \rhos(t)/\D t=-\frac i\hbar\com{\hat H}{ \rhos(t)}+\mathcal L[\rhos(t)]$, where $\hat H$ describes the free evolution of the system and
\bq\label{csl-eq}
\mathcal L[\rhos(t)]=-\frac{ \lcsl}{2\rC^3\pi^{3/2}m_0^2}\int \D{\bf z}\com{\hat M({\bf z})}{\com{\hat M({\bf z})}{\rhos(t)}},
\eq
governs the CSL effect on the system. $\hat M({\bf z})$ is defined as follows:
\bq\label{Mz-eq}
\hat M({\bf z})=m_0\sum_n \exp\left({-\tfrac{({\bf z}-\hat {\bf q}_n)^2}{2\rC^2}}\right),
\eq 
where $m_0$ is a reference mass chosen equal to the mass of a nucleon, the sum $\sum_n$ runs over all  nucleons of the system and $\hat \q_n$ is the position operator of the $n$-th nucleon. When the spread of the center of mass wavefunction is much smaller than $\rC$, which is typical of all situations we are interested in, we can Taylor expand to  second order in $\hat \q_n$ \cite{Carlesso:2016aa}, and rewrite Eq.~\eqref{csl-eq} as 
\bq\label{eqdiffus}
\mathcal L[\rhos(t)]=-\tfrac{1}{2}\sum_{i,j=x,y,z}\eta_{ij}\com{\hat q_{i}}{\com{\hat q_{j}}{ \rhos(t)}},
\eq
where {$\hat q_{i}$ is the center of mass position operator along the $i$-th direction and}
\bq\label{eqeta}
\eta_{ij}=\frac{\lambda \rC^3}{\pi^{3/2}m_0^2}\int\Dd{3}\k\, e^{-\rC^2k^2}k_ik_j|\tilde\mu(\k)|^2,
\eq
with $\tilde \mu(\k)$  the Fourier transform of the mass density of the system. 

Eq.~\eqref{eqdiffus} describes a diffusive dynamics, quantified by the CSL-induced diffusion constants $\eta_{ij}$, which can be best measured via optomechanical techniques~\cite{Bahrami:2014aa,Nimmrichter:2014aa,Diosi:2015ab}. 
In a typical experimental setup, the position of a mechanical resonator is accurately monitored and the force acting on it is determined; this is for instance the case of cantilever experiments~\cite{Vinante:2016aa,Vinante:2017aa} or gravitational wave detectors~\cite{Carlesso:2016aa,Helou:2017aa}.  In such a setup, diffusion is conveniently quantified by the Density Noise Spectrum (DNS) of the resonator's position, {which reads
\bq
\DNS=\frac12\int_{-\infty}^{+\infty} \D \tau \,e^{-i \omega \tau}{\mathbb E} [\braket{\{\delta\hat q_z(t), \delta \hat q_z(t+\tau)\}}],
\eq
where $\delta\hat q_z(t)=\hat q_z(t)-q_\text{\tiny ss}$ denotes the fluctuations in position along the $z$ direction, the measurement direction, with respect to the steady state position $q_\text{\tiny ss}$. The DNS is the quantity measured in the experiment and it quantifies the motion of system and its diffusive dynamics.}
Under the effect of thermal fluctuations and the CSL diffusion the DNS takes the form~\cite{carlesso:2018ab}:
\bq\label{dns}
\DNS=\frac{ 2M\gammam\kb T +\Scsl}{M^2[(\omega_{0}^2-\omega^2)^2+\gammam^2\omega^2]},
\eq
where $M$, $\omega_0$ and $\gammam$ are respectively the mass, the resonance frequency and the damping of the resonator, and $T$ is the temperature of the thermal noise ($\kb$ is Boltzmann constant). CSL contributes to the DNS as a temperature independent force noise equal to $\Scsl=\hbar^2\eta$, where $\eta=\eta_{zz}$ is the CSL diffusion constant along the $z$-direction, the direction of measurement. 

Eq.~\eqref{dns} shows that in order to increase the relative strength of the CSL effect with respect to the thermal noise, one has two options: one either  minimizes the thermal force noise $\mathcal S_{\text{\tiny th}}=2M\gammam\kb T$, which requires low temperatures and/or low damping regimes, or  maximizes the CSL force noise, i.e.~the diffusion constant $\eta$. 

Some of the strongest CSL bounds have been set by mechanical experiments, which were designed for ultralow thermal noise. For  experiments with cantilevers, this is achieved by operating at millikelvin temperature, for macroscopic experiments such as gravitational wave detectors, the key ingredient is the operation at very low frequency, where the mechanical damping can be strongly reduced. Further decrease of temperatures and/or low damping requires demanding technological improvements.

Here, we are interested in the other option: to explore possible ways to enhance the CSL diffusion by optimizing the shape and the mass density distribution of the test mass.  In a  cantilever experiment,  the 
damping constant $\gammam$ is mainly defined by the cantilever stiffness and the value of the attached mass, independently of its shape. Thus, at fixed mass, the shape plays a role only in defining $\eta$. {Quantitative calculations (see Appendix \ref{App.levitated}) show that the cuboidal geometry is preferable over the spherical, since is the one that shows the strongest CSL diffusion. }Similar results can be obtained also for a cylindrical geometry, once the ratio between the base length and the height of the system is properly chosen. For the sake of simplicity in the following analysis we will focus on the cuboidal geometry. 

\begin{figure*}[th!]
\centering
\includegraphics[width=0.9\linewidth]{{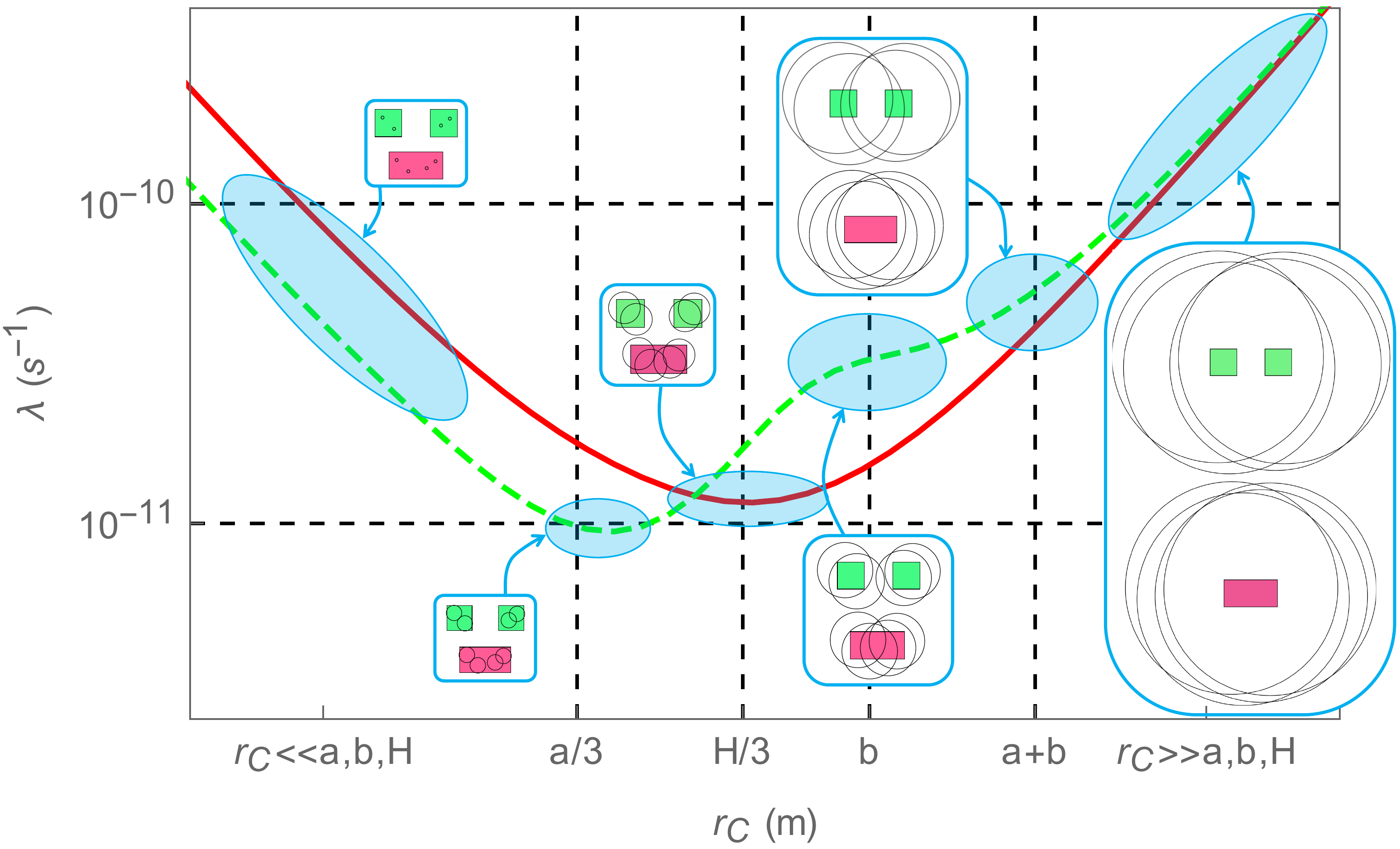}}
\caption{\label{plottotab}{ Hypothetical bounds on the CSL parameters obtained by monitoring the center of mass motion of an harmonically trapped system of mass $M$. Two configuration are considered: a cuboid of side $H$ (red) and two cuboids of side $a=H/2$ separated by a distance $b$ which is supposed to be constant (green dashed). The base area is the same for both configurations.
 The circles represent Gaussians of variance $\rC^2$ inside which the CSL noise acts coherently. Due to the two different geometrical configurations (see Appendix \ref{App2}) of the mass, the bounds become stronger or weaker depending on the value of $\rC$. In particular, for relatively small values of $\rC$ a non-uniform mass density makes the bound stronger (roughly by a factor of 2 for the considered configurations). This feature of the collapse mechanism is at the heart of the amplification effect of the multilayer structure which is discussed in the paper.
 }}
\end{figure*}

Preliminary heuristic considerations can be done by looking at the characteristic profile of the upper bounds inferred from non-interferometric experiments \cite{Vinante:2016aa,Carlesso:2016aa,Helou:2017aa,Vinante:2017aa}, c.f.~light orange lines in Fig.~\ref{fig1}. Such a profile can be  understood by looking at Fig.~\ref{plottotab}: for a single mass, the CSL effect (as well as the bound on $\lcsl$) is strongest  when $\rC\sim H/3$, where $H$ defines the mass dimension. Conversely, for $\rC\ll H$ or $\rC\gg H$ the effect is weakened by its incoherent or unfocused action respectively. 

In the following, we will quantify such a profile for a system composed by $N$ masses, modeled as harmonic oscillators. The action of the CSL noise on each mass can be described in terms of the Langevin equations \cite{Carlesso:2016aa}: {
\bqali\label{langevinalpha}
\frac{\D \hat \q_\alpha}{\D t}&=\frac{\hat \p_\alpha}{m_\alpha},\\
\frac{\D \hat \p_\alpha}{\D t}&=-m_\alpha\omega_\alpha^2\hat \q_\alpha-\gamma_\alpha\hat\p_\alpha+\hat{\bm \xi}_\alpha+\F_\alpha,
\eqali}
where $\omega_\alpha$, $\gamma_\alpha$ and $m_\alpha$ are respectively the frequency, the damping constant and the mass of the $\alpha$-th mass. {$\hat{\bm \xi}_\alpha$ }and $\F_\alpha$ are the surrounding environmental and the CSL stochastic forces, whose action leads to thermal and non-thermal diffusions, respectively. 
{Going into the details,} the correlations of the CSL forces depend on the distance between the masses. In the limit of validity of Eq.~\eqref{eqdiffus}, the Fourier transform of $\F_\alpha$ becomes \cite{Carlesso:2016aa}
\bq
\tilde \F_\alpha=\frac{i \hbar \sqrt{\lcsl}\rC^{3/2}}{(4\pi^3)^{3/4}m_0}\!\int\D\z\,\tilde w(\z,\omega)\!\int\D\k\,\tilde \mu_\alpha(\k)e^{-\tfrac{\k^2r_C^2}{2}-i\k\cdot\z}\,\k,
\eq
where $\tilde \mu_\alpha(\k)$ and $\tilde w(\z,\omega)$ are respectively the Fourier transform of the mass density $\mu_\alpha(\x)$ of the $\alpha$-th mass and of a white noise. For the latter, it holds: $\braket{\tilde w(\z,\omega)}=0$ and $\braket{\tilde w(\z,\omega)\tilde w(\z',\Omega)}=2\pi\delta(\omega+\Omega)\delta^{(3)}(\z-\z')$. Consequently, the correlations read:{
\begin{multline}
\label{corrF}
\braket{{\tilde F_{\alpha,i}(\omega)}{\tilde F_{\beta,j}(\Omega)}}\\=\frac{2\hbar^2\lcsl \rC^3\delta(\omega+\Omega)}{\sqrt{\pi}m_0^2}\int\D\k\,\tilde\mu_\alpha(\k)\tilde\mu_\beta^*(\k)e^{-\k^2\rC^2}k_ik_j,
\end{multline}}
which reduces to $\braket{{\tilde F_{i}(\omega)}{\tilde F_{j}(\Omega)}}=2\pi\hbar^2\delta(\omega+\Omega)\eta_{ij}$ for $N=1$, with $\eta_{ij}$ defined in Eq.~\eqref{eqeta}.

We are interested in the motion of the center of mass of the system, whose dynamical equation can be derived from Eq.~\eqref{langevinalpha}: 
\bqali
\frac{\D \hat \q_\text{\tiny cm}}{\D t}&=\frac{\hat \p_\text{\tiny cm}}{M},\\
\frac{\D \hat \p_\text{\tiny cm}}{\D t}&=-M\omega_0^2\hat \q_\text{\tiny cm}-\gammam\hat\p_\text{\tiny cm}+\hat{\bm \xi}_\text{\tiny cm}+\F_\text{\tiny cm},
\eqali
where $M=\sum_\alpha m_\alpha$, and we set $\omega_\alpha=\omega_0$ and $\gamma_\alpha=\gammam$. This is the case when the masses are clamped together and attached to a cantilever, thus they move together at the frequency $\omega_0=\sqrt{k/M}$ where $k$ is the cantilever stiffness, while the damping $\gammam$ will be typically determined by cantilever bending losses. {We also defined ${\F}_\text{\tiny cm}=\sum_\alpha{ \F}_\alpha$ and $\hat{\bm \xi}_\text{\tiny cm}=\sum_\alpha\hat{\bm \xi}_\alpha$. The correlations of ${\F}_\text{\tiny cm}$ can be derived from Eq.~\eqref{corrF}. The environmental noise is preponderately due to the dissipation of the cantilever spring and its correlations read $\tfrac12\braket{\acom{\hat\xi_{\text{\tiny cm},i}(t)}{\hat\xi_{\text{\tiny cm},j}(s)}}=2M\gamma \kb T\delta_{i,j}\delta(t-s)$ (with $i,j=x,y,z$), which depend on the total mass of the system and the damping of the cantilever only \cite{Vinante:2016aa,Vinante:2017aa}. From the form of these correlations, one can derive} the thermal and non-thermal (CSL) contributions, whose form is $\mathcal S_{A}=\int\D\Omega\,{\braket{\acom{\tilde{A}(\omega)}{\tilde{A}(\Omega)}}}/{4\pi}$, to the DNS, which was introduced in Eq.~\eqref{dns}. By applying the correlation rules for $\hat{\bm \xi}_\text{\tiny cm}$ and ${ \F}_\alpha$ previously outlined, we end up with:
\bqali\label{SNmass}
\mathcal S_{\text{\tiny th}}&=2M\gammam\kb T,
\\
\Scsl&=\frac{\hbar^2\lcsl \rC^3}{\pi^{3/2}m_0^2}\int\D\k\,\sum_{\alpha,\beta}\left(\tilde\mu_\alpha(\k)\tilde \mu_\beta^*(\k)\right)e^{-\k^2r_C^2}k_z^2,
\eqali
where we focused once again on the motion along the $z$-direction, which is assumed to be the direction of measurement. If $N=1$, these relations correspond to those entering Eq.~\eqref{dns}.

{We note that $\mathcal S_{\text{\tiny th}}$ is proportional to the mass $M$. Indeed, the main contribution to the thermal noise comes from the coupling to the cantilever spring, thus depending only on the total mass $M$ and $\gammam$ \cite{Vinante:2016aa,Vinante:2017aa}. Consequently, the thermal noise does not change if the system is composed by one or many layers for a fixed value of the mass. 
}
On the other hand, {the CSL force acts directly on the mass layers and} $\Scsl$ is a sum of $N^2$ contributions: $N$ contributions are due to the self-correlation of a single  mass; $N(N-1)$ are due to the cross-correlation terms. While the former are positive by definition, the latter do not have a definite sign, and depend on the distance $d_{\alpha,\beta}$ between the $\alpha$-th and $\beta$-th mass. Indeed, by considering only two masses, if $\rC\ll d_{\alpha,\beta}$, the forces acting on the two masses are uncorrelated, hence the corresponding cross-correlation term vanishes. If $\rC\gtrsim d_{\alpha,\beta}$, the two forces contribute coherently to the center of mass diffusion: this is the situation that maximizes the CSL effect. If $\rC\gg d_{\alpha,\beta}$, the main contribution to the integral in Eq.~\eqref{SNmass} comes from $|\k|<1/d_{\alpha,\beta}$; the rest is suppressed due to the Gaussian weight and consequently the global CSL effect does not benefit from it. This analysis is summarized in Fig.~\ref{plottotab} for two masses.

A first example of this analysis is reported in Fig.~S5 of \cite{Vinante:2016aa}, where the \emph{mixed} term diminishes the self-correlated contributions to $\Scsl$ for $\rC\lesssim10^{-6}\,$m, a value for which both the noise acting on the cantilever and that acting on the sphere coherently contribute to the CSL diffusion of the center of mass of the system. For $\rC\gtrsim10^{-6}\,$m the \emph{mixed} contribution to $\Scsl$ is positive and for $\rC\gg10^{-6}\,$m it goes to zero.
\begin{figure}[t!]
\centering
\includegraphics[width=0.6\linewidth]{{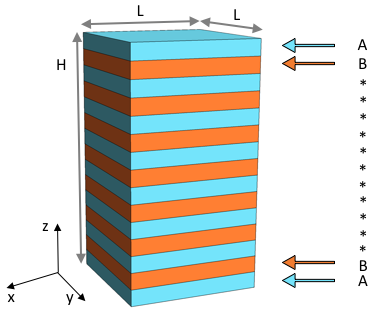}}
\caption{\label{figuracuboide}Graphical representation of the $2N_\text{\tiny lay}+1$ layered cuboid. The two materials A and B are represented with two different colours, respectively cyan and orange.}
\end{figure}
We will now discuss a concrete application of the model discussed in the previous section. { Let us consider a test mass consisting of a cuboid of base $L\times L$ and height $H$ made of $(2\Nlay+1)$ layers, which are parallel to the base and orthogonal to the $z$-axes.} These layers are made of two different materials, respectively $\Nlay+1$ of material A (density $\muA$) of thickness $a$ and $\Nlay$ of material B (density $\muB$) of thickness $b$, alternatively disposed one on top of the other. In order to maximize the contribution to $\eta$, we choose the layers labeled by A, whose number exceed the B layers by one, to be the heavier ones [cf.~Fig.~\ref{figuracuboide}]. 

Thus, by carefully tuning the dimension $L$ of the single mass and the distance $d$ between the masses, one can explore different CSL parameter regions even though the value of $\Scsl$ does not change.

\section{Multilayer approach}

The test mass is supposed to be attached to a cantilever. Specifically, we take as a reference the experiment described in Ref.~\cite{Vinante:2017aa}. Here, the resonant frequency is $\omega_0/ 2 \pi= 8174$ Hz, the spring constant $k=0.40$ N/m and the test mass is a NdFeB sphere with density $\mu_s=7430$\,kg/m$^3$ and radius $R=15.5\,\mu$m. Under these conditions the measured residual force noise acting on the cantilever after subtracting the thermal noise is $S_F=2.0$ aN$^2$/Hz. This value corresponds to an excess noise of unknown origin, compatible with CSL. Here, we want to probe the values of $\lambda$ and $\rC$ that correspond to such a value.

While keeping all other experimental parameters fixed, we replace now the NdFeB sphere with a layered cuboid with the same mass, and variable geometry ($L$, $a$, $b$ and $\Nlay$). By taking the same mass we keep also the same resonant frequency, so to guarantee a fair comparison with the experiment in Ref.~\cite{Vinante:2017aa}. We choose the densities of the two materials equal to $\muA=16.0 \times 10^3$\,kg/m$^3$ and $\muB=2.2 \times 10^3$\,kg/m$^3$, which correspond respectively to CoPt, a heavy ferromagnetic material required for the SQUID detection, and SiO$_2$. 
SiO$_2$ is one of the most common materials, it is easy to fabricate and comparatively light. CoPt is one of the heaviest ferromagnetic materials, and is chosen here to enable SQUID detection in absence of a magnetic sphere as in Ref.~\cite{Vinante:2017aa}. If the latter condition is not required, better choices for the heavy material are for instance Au and W, whose densities are almost the same $\muA=19.41\times10^3\,$kg/m$^3$. 
Given the measured value of the residual force noise, we compute the upper bounds on the CSL parameters for different values of the cuboid parameters. The Fourier transform of cuboidal mass density is given by
\begin{figure}[t!]
\includegraphics[width=\linewidth]{{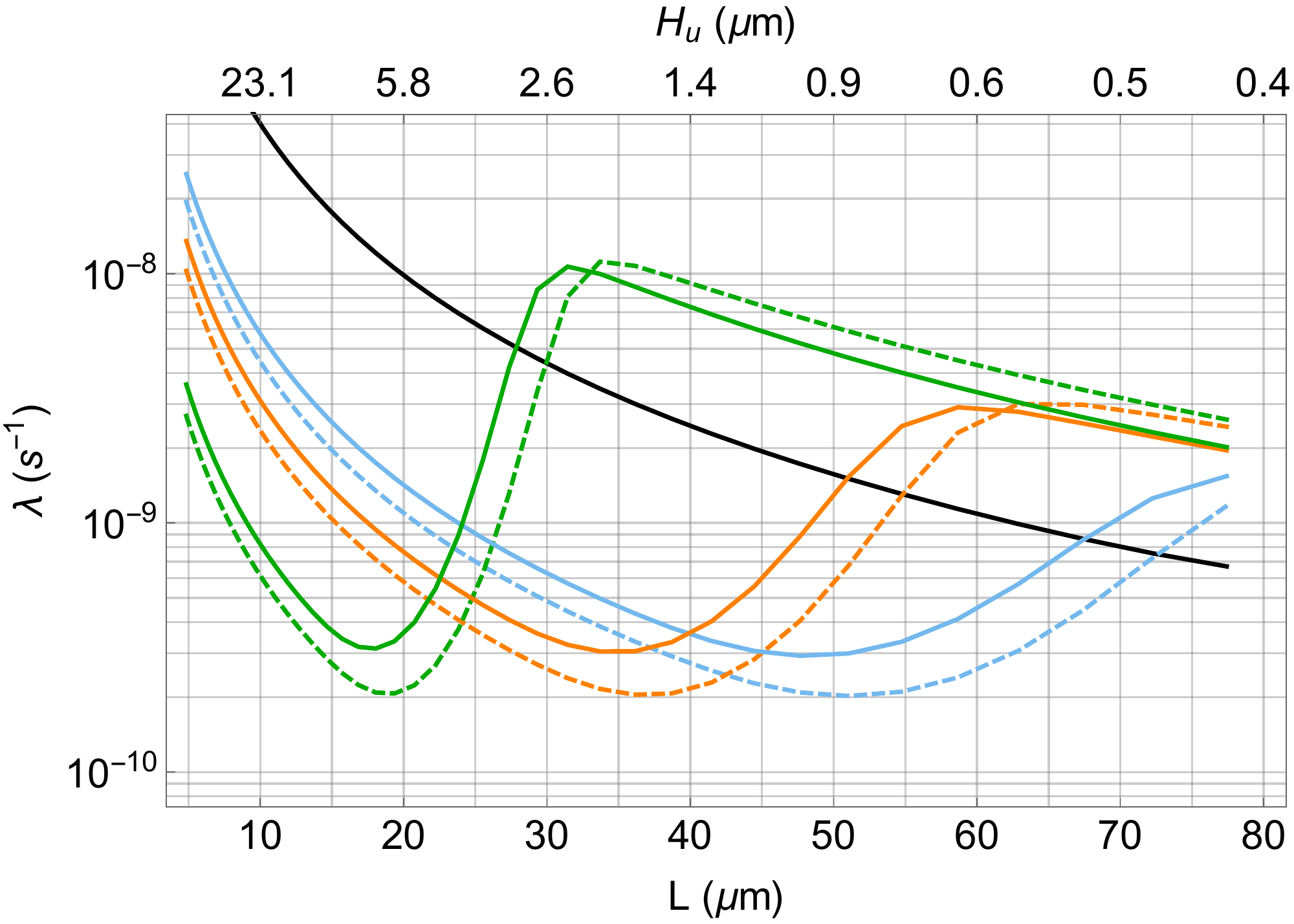}}
\caption{Comparison of the hypothetical upper bounds on $\lambda$ for different configurations of the test mass for $\rC=10^{-7}\,$m for different values of $\Nlay$ with $a=b$. The uniform case (black line) is compared with the multilayer approach for $\Nlay=8$ (blue lines), $\Nlay=16$ (orange lines) and $\Nlay=64$ (green lines). Two cases are checked: $\muA=16.0 \times 10^3$\,kg/m$^3$ and $\muB=2.2 \times 10^3$\,kg/m$^3$ (continuous lines) and $\muA=16.0 \times 10^3$\,kg/m$^3$ and $\muB=0$ (dashed lines). { The mass is held fixed to $M=1.2\times10^{-10}\,$kg. }The top horizontal axis indicates $H$ in the uniform case.\label{plotcomp1}}
\end{figure}
\bq\label{mutildecylinder}
\tilde \mu(\k)=\frac{4}{k_xk_y}\sin(\tfrac{k_xL}{2})\sin(\tfrac{k_yL}{2})
\tilde \mu_z(k_z),
\eq
where
\begin{multline}\label{eqdensity}
\tilde \mu_z(k_z)=\frac{2}{k_z\sin\left(\tfrac{k_z(a+b)}{2}\right)}\cdot\\
\cdot\left[\muA\sin\left(	\tfrac{k_za}{2}\right)\sin\left(\tfrac{k_z(H+b)}{2}\right)+\right.\\
\left.+\muB\sin\left(	\tfrac{k_zb}{2}\right)\sin\left(\tfrac{k_z(H-a)}{2}\right)	\right],
\end{multline}
with $H=(\Nlay+1)a+\Nlay b$. In the particular case $a=b$, the latter expression reduces to 
\bqali\label{mutildecylinderz}
\tilde \mu_z(k_z)=
\frac{\mu_\text{\tiny A}\sin\left((\Nlay+1)k_za\right)+\mu_\text{\tiny B}\sin\left(\Nlay k_za\right)}{k_z\cos(\tfrac{k_za}{2})e^{-ik_z(H/2+a)}}.
\eqali
Combining Eq.~\eqref{mutildecylinder} and Eq.~\eqref{eqeta}, we obtain the CSL diffusion constant
\bq
\eta=\frac{16 \rC^5\lambda}{m_0^2\sqrt{\pi}}\left[1-e^{-\tfrac{L^2}{4\rC^2}}-\frac{L\sqrt{\pi}}{2\rC}\erf\left(\tfrac{L}{2\rC}\right)\right]^2
\cdot \mathcal I_z,
\eq
where 
\bq\label{defIz}
\mathcal I_z=\int\D k_z\,e^{-\rC^2k_z^2}k_z^2|\tilde\mu_z(k_z)|^2.
\eq
The latter must be in general computed numerically. For the special case of $\Nlay=0$, we obtain the standard expression \cite{Nimmrichter:2014aa}
\bq\label{defIzN0}
\mathcal I_z=\frac{2\sqrt{\pi}\mu_\text{\tiny A}^2}{\rC}\left(1-e^{-\tfrac{H^2}{4\rC^2}}\right),
\eq
where here $H=a$.

\begin{figure}[t!]
\includegraphics[width=1\linewidth]{{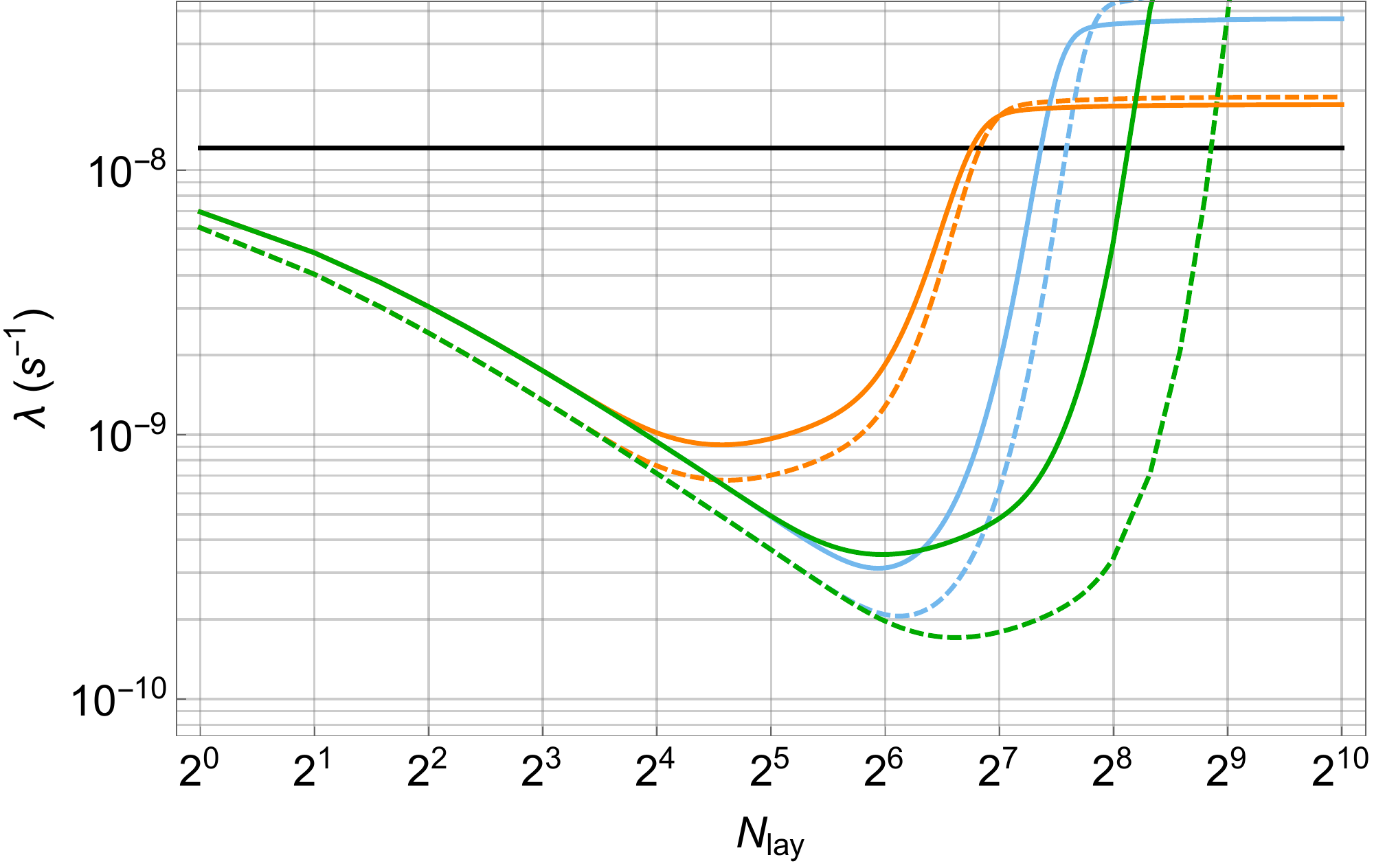}}
\includegraphics[width=1\linewidth]{{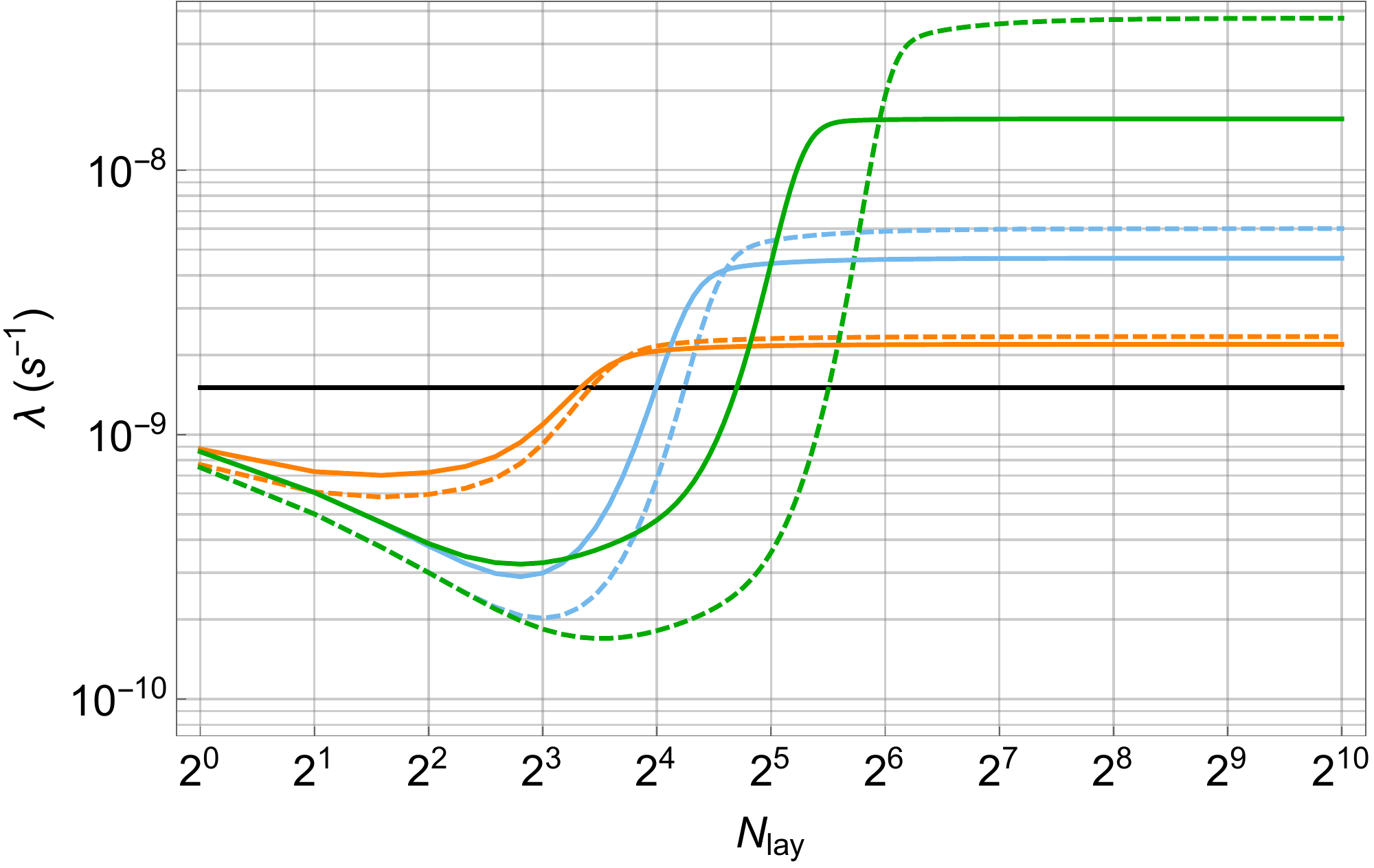}}
\caption{Comparison of the hypothetical upper bounds on $\lambda$ for $\rC=10^{-7}\,$m for different values of $\Nlay$ and $\epsilon=b/a$ with $L=18\,\mu$m (top panel) and $L=50\,\mu$m (bottom panel). The uniform case (black line) is compared with the multilayer approach for $\epsilon=1/4$ (orange lines), $\epsilon=1$ (blue lines) and $\epsilon=4$ (green lines). { The densities are fixed at $\muA=16.0 \times 10^3$\,kg/m$^3$ and $\muB=2.2 \times 10^3$\,kg/m$^3$. The extreme case with $\muB=0$ is reported with dashed lines. The mass is held fixed to $M=1.2\times10^{-10}\,$kg.} \label{plotcomp2}}
\end{figure}

\begin{figure*}[ht!]
\centering
\includegraphics[width=0.5\linewidth]{{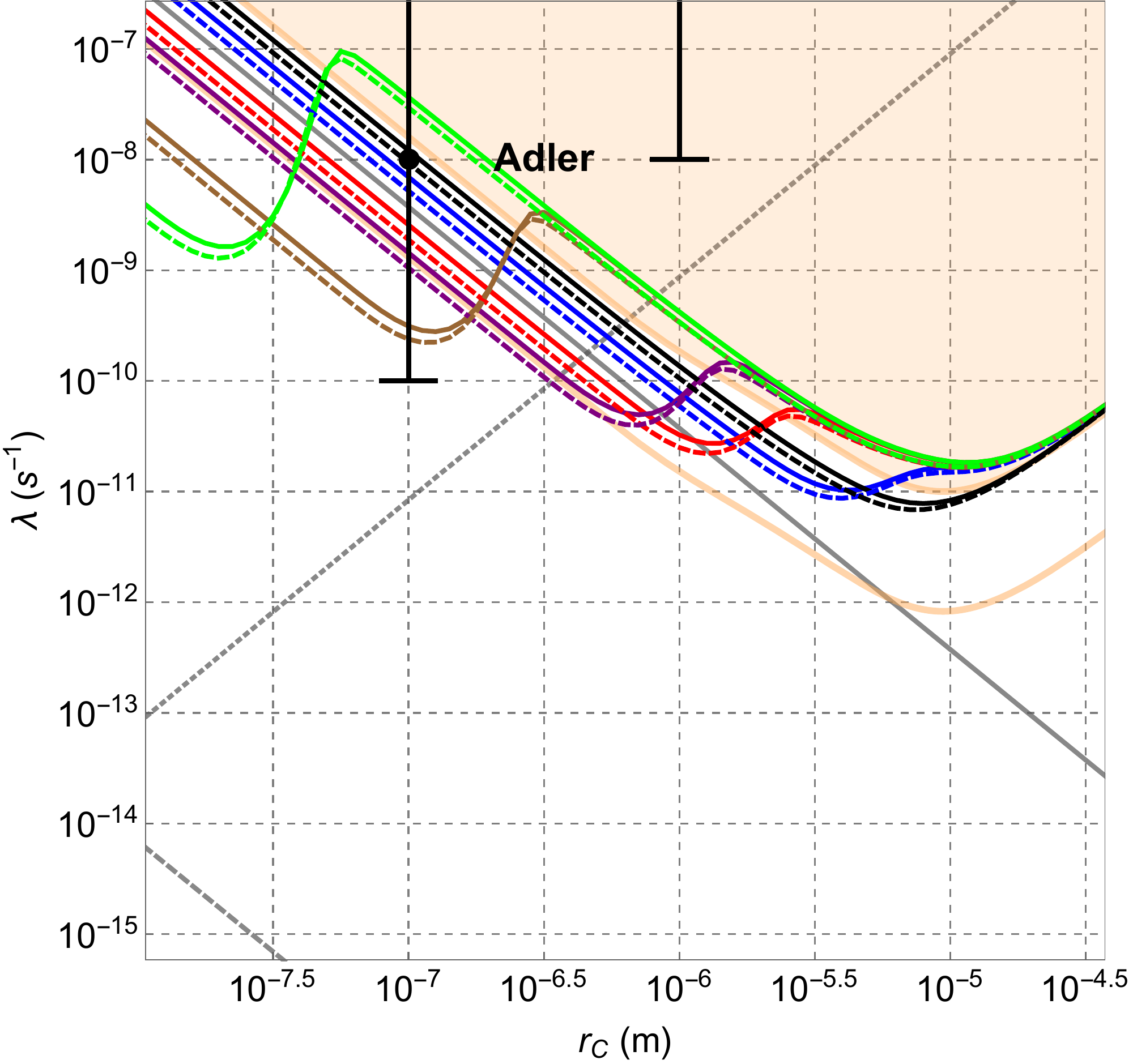}}\includegraphics[width=0.5\linewidth]{{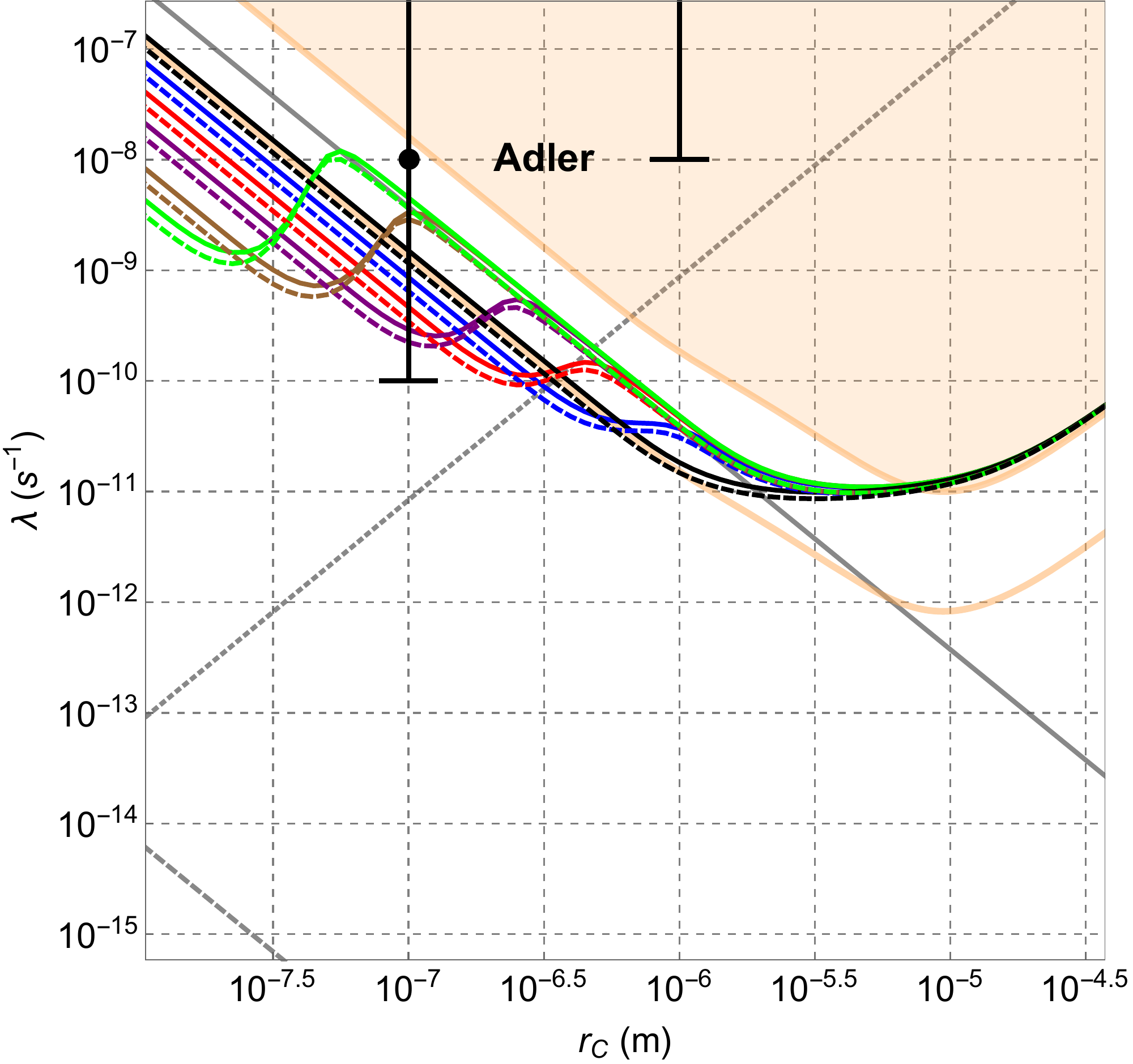}}
\caption{\label{fig1} Hypothetical upper bounds from multilayer test mass on the same cantilever of Ref.~\cite{Vinante:2017aa}. The considered system is a cuboid of base side $L=18.058\,\mu$m (left panel) and $L=51.077\,\mu$m (right panel)  {at fixed mass $M=1.2\times10^{-10}\,$kg}. Left panel: black, blue, red, purple, brown and green  {continuous }lines refer to $\Nlay=0,\,1,\,5,\,10,\,61,\,360$ respectively. Right panel: black, blue, red, purple, brown and green  { continuous} lines refer to $\Nlay=0,\,1,\,3,\,7,\,20,\,40$ respectively.
Light orange lines (and the corresponding shaded area) represent respectively the excess noise measured in Ref.~\cite{Vinante:2017aa} and, if the latter would not result as a CSL effect, the upper bound from cantilever experiment \cite{Vinante:2017aa}.  {The densities are fixed at $\muA=16.0 \times 10^3$\,kg/m$^3$ and $\muB=2.2 \times 10^3$\,kg/m$^3$.  The extreme case with $\muB=0$ is reported with dashed lines. For comparison, we report with grey lines other significant experimental upper bounds: X-ray spontaneous emission (dotted line) \cite{Curceanu:2015aa}, LISA Pathfinder (continuous line)\cite{Carlesso:2016aa} and theoretical lower bound (dashed line) \cite{Toros:2017aa}.} Other weaker experimental bounds are not reported \cite{Laloe:2014aa,Belli:2016aa,Bilardello:2016aa,Toros:2017aa}.}
\end{figure*}
We start our numerical analysis by noting that, once the value of the mass {and the material densities are fixed,} $\eta$ depends on three parameters of the system: the base side $L$ of the cuboid, the number of layers $(2\Nlay+1)$ and the ratio between the thickness of the two materials $\epsilon=b/a$.\\
\indent We take the value $\rC=10^{-7}\,$m as a reference, and by fixing $\epsilon=1$ and $\Nlay=1,\,16$ and 64, we compute the hypothetical bounds obtained by varying $L$. Fig.~\ref{plotcomp1} compares the bounds from the uniform case (black line) with the ones obtained by using the multilayer approach (colored lines), with $\Nlay=1$ (blue lines), $\Nlay=16$ (orange lines) and $\Nlay=64$ (green lines). To underline the importance of the density difference between the two materials used, Fig.~\ref{plotcomp1} shows the bounds obtained using $\muA=16.0 \times 10^3$\,kg/m$^3$ and $\muB=2.2 \times 10^3$\,kg/m$^3$ (continuous lines) and the extreme case with $\muA=16.0 \times 10^3$\,kg/m$^3$ and $\muB=0$ (dashed lines). Remarkably, a large density difference enhances the CSL signal.

As Fig.~\ref{plotcomp1} shows, by using the multilayer approach, one can gain almost two orders of magnitude in bounding $\lambda$ with respect to the uniform case, c.f.~$L\sim20\,\mu$m and $\Nlay=64$. The choice of the range of possible values of $L$ is constrained by experimental considerations. The test mass should be accommodated on the cantilever, so $L$ is limited by the cantilever width. In the opposite high aspect ratio limit $H/L\gg1$, the test mass becomes a thin pillar and it cannot be treated as a simple inertial mass anymore. A good compromise is a value of $L$ which is comparable with $H$, for instance we consider for the further analyses $L=18\,\mu$m. For a comparison, we consider also a bigger, but still worthwhile, value $L=50\,\mu$m.\\

As the second step of our numerical investigation, we fix the value of the side length $L$ to the values defined above and vary $\Nlay$ and $\epsilon$.
Fig.~\ref{plotcomp2} compares the bound given by the uniform mass (black line) with those of the multilayer approach for different values of $\Nlay$ with $\epsilon=1/4$ (continuous orange line), $\epsilon=1$ (continuous blue line) and $\epsilon=4$ (continuous green line). Again, we also studied the case of $\muB=0$, whose data are reported with the corresponding dashed lines.

\begin{table}[t!]
\begin{tabular}{|c|c|c|c|c|}
\hline\hline
$L\, [\mu$m] &\ $H\, [\mu$m]&\ $\Nlay$& $a=b\, [\mu$m]&$\lambda\, [$s$^{-1}$] \\ 
 \hline\hline
{ 18}& $39$ &$61$&$0.32$ &$3.1\times10^{-10}$ \\ 
{ 50}& $4.6$ &$7$&$0.31$ &$2.9\times10^{-10}$ \\ 
\hline\hline
\end{tabular}
\caption{\label{tablebestN} Parameters of the test mass that maximize the bound on $\lambda$ from the analysis shown in Fig.~\ref{plotcomp2}.}
\end{table}
Fig.~\ref{plotcomp2} shows that the best configuration is given by $\Nlay=61$ with $\epsilon=1$ for $L=18\,\mu$m and by $\Nlay=7$ with $\epsilon=1$ for $L=50\,\mu$m. The corresponding values of $H$, $a$, $b$ and the bound on $\lambda$ for $\rC=10^{-7}\,$m are reported in Table \ref{tablebestN}. It is worthwhile to notice that, although the dimensions of the proposed test masses are well different, the value of $a$ (and equivalently $b$) is almost identical in the two configurations that maximize the bound on $\lambda$. The optimal value of $a\sim b$ is of the order of $\rC$, which is in agreement with the heuristic argument discussed in Fig.~\ref{plottotab}. \\

As the last step of the analysis, we compute the hypothetical bounds in the CSL parameters space ($\rC$ vs $\lambda$) for the configurations reported in Table \ref{tablebestN}. These are reported in Fig.~\ref{fig1} for different values of $\Nlay$. It is clear that with the multilayer approach one can strongly improve the bound on $\lambda$ by one or two orders of magnitude, depending on the side length.\\

\begin{figure}[t!]
\includegraphics[width=1\linewidth]{{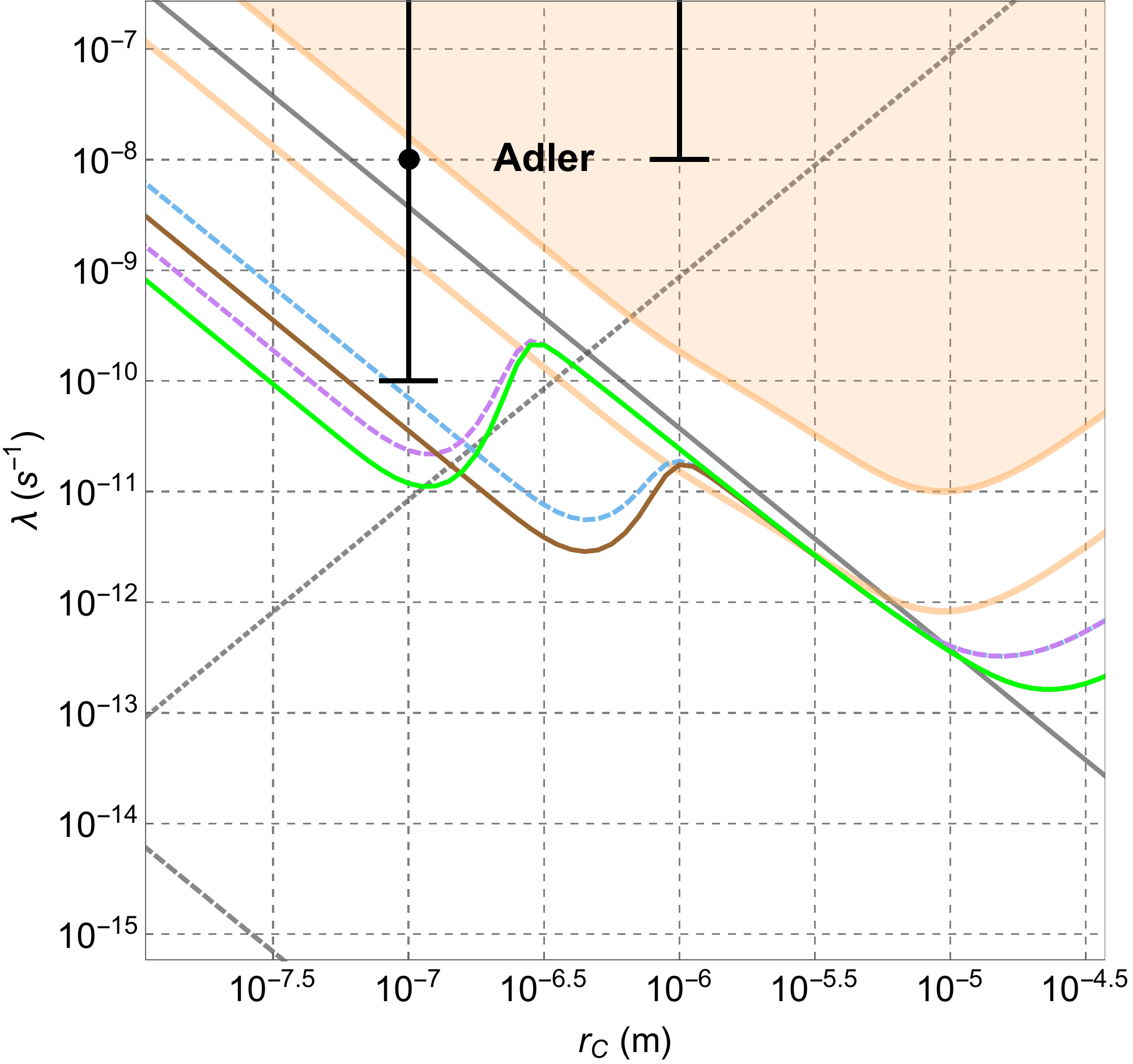}}
\caption{Hypothetical upper bounds from multilayer method with a bigger test mass. The  system here considered is a cuboid of base side $L=60\,\mu$m. Dashed purple and blue lines correspond to $M_1=1.16 \times 10^{-9}$\,kg with $\Nlay=48$ and $\Nlay=12$ respectively. Green and brown lines correspond to $M_2=2.32 \times 10^{-9}$\,kg with $\Nlay=98$ and $\Nlay=25$ respectively.
The other lines and the colored region refer to ranges of parameters of CSL, which are already excluded by other experimental data, as described in Fig.~\ref{fig1}.\label{plotbiggermass}}
\end{figure}

Since the CSL effect scales with the total mass of the mechanical oscillator, it is worth to extend the analysis to larger masses. Specifically, we consider  $M_1=1.16 \times 10^{-9}$\,kg and $M_2=2.32 \times 10^{-9}$\,kg, which are respectively 10 and 20 times larger than the mass previously considered. By keeping $L=60\,\mu$m and $\epsilon =1$, we chose $\Nlay$ such that the second minimum in $\lambda$ appears near $\rC=10^{-7}\,$m. This corresponds to having $a=b\simeq 0.3\,\mu$m [cf.~Tab.~\ref{tablebestN}] and taking $\Nlay=48$ ($H\simeq29\,\mu$m) and $\Nlay=98$ ($H\simeq59\,\mu$m) for the two cases respectively. We consider also the case with $\Nlay=12$ ($a\simeq1.2\,\mu$m, $H\simeq29\,\mu$m) and $\Nlay=25$ ($a\simeq0.6\,\mu$m, $H\simeq29\,\mu$m) respectively as a comparison. We note that also with this increased size, the test mass would still fit on the cantilever of Ref.~\cite{Vinante:2017aa}. Fig.~\ref{plotbiggermass} shows the corresponding bounds assuming that the value of the measured noise remains the same as in~\cite{Vinante:2017aa}. This is a stronger assumption with respect to the previous analysis, since also the resonant frequencies will change according to $\omega_i=\sqrt{k/M_i}$, which gives $\omega_1/(2\pi)=2584\,$Hz and $\omega_2/(2\pi)=1828\,$Hz respectively. With this assumption, the multilayer configuration for a mass equal to $M_2$ with 48 layers is able to test the CSL model almost down to $\lambda=10^{-11}\,$s$^{-1}$. Thus, this method can provide bounds comparable with those from the X-ray measurements, which, contrary to cantilever experiments, are  less robust against changes in the CSL noise~\cite{Nobakht:2018aa,Carlesso:2018aa}.

\section{Discussion}

The novel feature of a multilayer cuboidal resonator is the appearance of a second minimum in the curve defining the upper bound. According to Fig.~\ref{plottotab}, while the main minimum corresponds to $\rC\sim H/3$, the new minumum appears at $\rC\sim a,b$ and it moves to smaller values of $\rC$ as $\Nlay$ increases. The reason for this behavior is the following: For small $\rC$ the single layer contributions add incoherently with the maximum effect when reaching $\rC\sim a,b$. For $\rC>a,b$ the cross-correlation between the layers interfere and the global diffusive action narrows until $\rC$ is of the order of the dimension of the system, when again the whole mass contributes coherently to the diffusive dynamics. As Fig.~\ref{fig1} shows, there is not an advantage of using a multilayer strategy for $\rC>a,b$.

The new hypothetical bounds are stronger than the bounds from the measured non-thermal excess noise reported in \cite{Vinante:2017aa}. Moreover they 
partially cover the orange highlighted region, which is the portion of CSL parameter space which results by attributing such an excess noise to standard sources.

Notably, the potential improvement would cover almost completely Adler's suggestion, $\lcsl=10^{-8\pm2}\,$s$^{-1}$ at $\rC=10^{-7}\,$m, using the value of the mass as in Ref.~\cite{Vinante:2017aa} [cf.~Fig.~\ref{fig1}].  
So far, Adler's values for the parameters have been ruled out only by two experiments. The first is the X-ray experiment~\cite{Curceanu:2015aa}, whose bound however may be evaded by a colored version of CSL \cite{Bassi:2010aa,Carlesso:2018aa}, with a frequency cutoff lower than $10^{18}$\,Hz, which is realistic. The second is the measurement of the crystal phonon excitations at low temperatures, however again the bound does not hold for colored extensions of the CSL model, and for a cutoff of the order of $10^{11}\,$Hz it vanishes \cite{Carlesso:2018aa}.
In both cases, an exclusion by a purely mechanical experiment would be much more significant since the dominant frequencies are much smaller. This is the case for the multilayer method applied to a larger mass [cf.~Fig.~\ref{plotbiggermass}].

One should also note that, differently from  previous  experiments, where for each value of $\rC$ one can at most infer a bound on $\lcsl$, the multilayer strategy  enables the possibility of identifying the value for $\rC$, if the presence of an excess noise were confirmed, by changing the geometry of the resonator. 

Finally, we underline that the hereby proposed scheme to enhance the CSL action can be easily implemented also in other type of mechanical resonators, as for example the one considered in \cite{Bahrami:2014aa,Nimmrichter:2014aa,Diosi:2015ab,Carlesso:2016aa,carlesso:2018ab,McMillen:2017aa,Schrinski:2017aa}.

\acknowledgments
\noindent  
The authors acknowledge support from the H2020 FET project TEQ (grant n.~766900).  AB acknowledges financial support from the University of Trieste (FRA 2016), INFN, the COST Action  QTSpace (CA15220), and hospitality from the IAS Princeton, where part of this work was carried out and partial financial support from FQXi.

%


\appendix

\section{CSL action on levitated systems}
\label{App.levitated}

\begin{figure*}[th!]
\centering
\includegraphics[width=0.5\linewidth]{{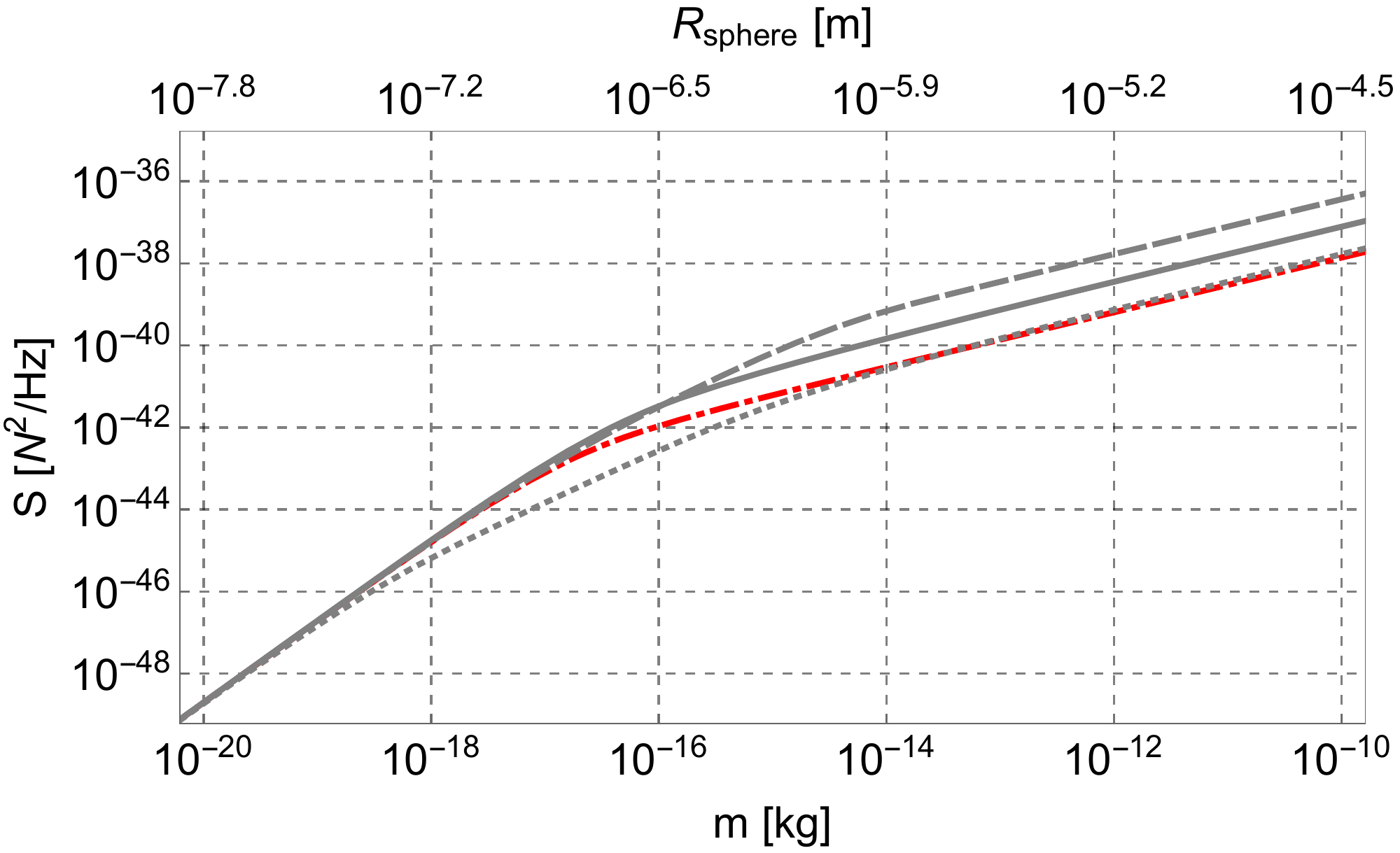}}\includegraphics[width=0.5\linewidth]{{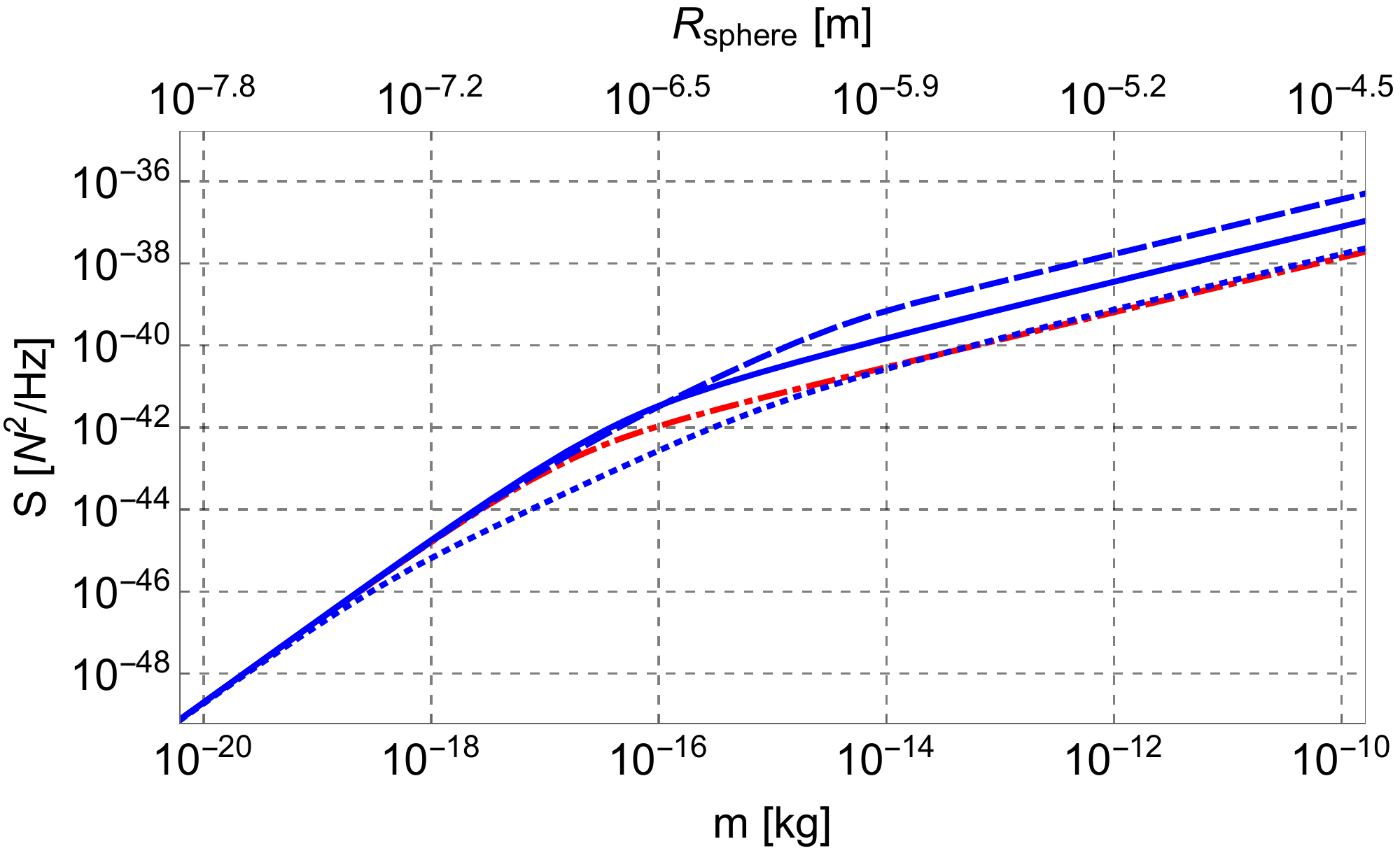}}
\caption{\label{thermalvsnonthermal} CSL contribution $\Scsl=\hbar^2\eta$ to the DNS as a function of the mass of the system; the density has been set equal to $\mu=2650$\,kg/m$^3$. [Top axis: for a better comparison, we report the value of the  radius of a sphere with given mass.] Left panel: spherical (red dot-dashed line) vs cuboidal (grey lines) geometry. Right panel: spherical (red dot-dashed line) vs cylindrical (blue lines) geometry. We considered three different aspect ratios for the cuboidal and for the cylindrical geometries: $L/H=0.1$ (dotted lines), $L/H=1$ (continuous lines) and $L/H=10$ (dashed lines). For the CSL parameters, we take as reference Adler's values: $\lcsl=10^{-8}$\,s$^{-1}$ and $\rC=10^{-7}$\,m.}
\end{figure*}

\begin{figure}[t!]
\centering
\includegraphics[width=\linewidth]{{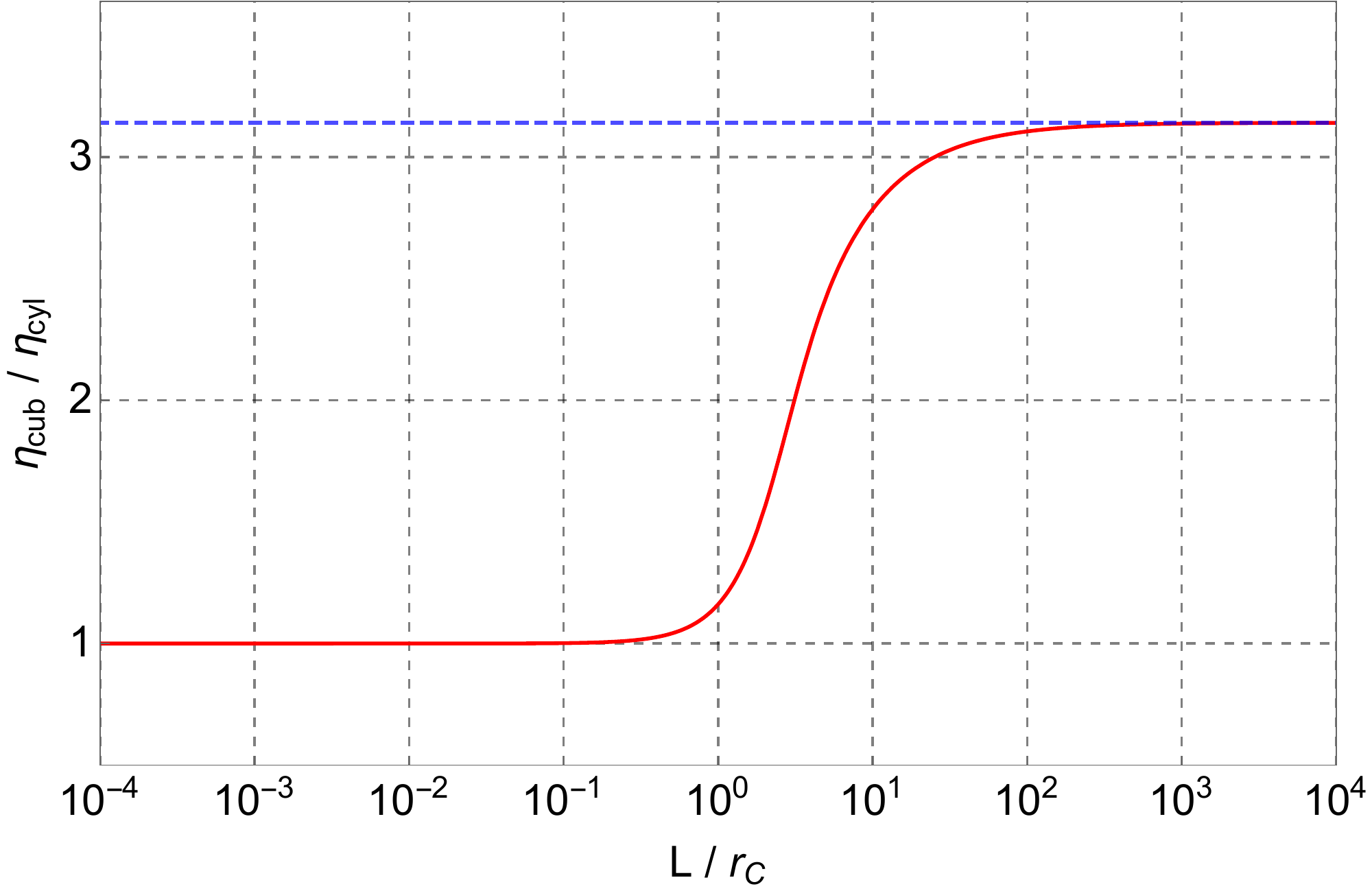}}
\caption{\label{cubvscyl} Comparison of the CSL diffusion rate for a cylinder $\eta_\text{\tiny cylinder}$ and for a cuboid $\eta_\text{\tiny cuboid}$, whose ratio depends only on $L/\rC${, independently from the mass and density of the system. The blue dashed line corresponds to the asymptotical value $\pi$.}}
\end{figure}

We compare the CSL contribution $\Scsl=\hbar^2\eta$ to the density noise spectrum $\DNS$ for three different cases: a sphere of radius $R$, a cuboid of lengths $(L,L,H)$, and a cylinder of radius $L$ and height $H$ (moving along the symmetry axis), all made of SiO$_2$ with density $\mu=2650$\,kg/m$^3$. The corresponding CSL contributions 
can be computed analytically \cite{Nimmrichter:2014aa}
\bqali
\eta_\text{\tiny (sphere)}&=\frac{3\lcsl m^2\rC^2}{m_0^2R^6}\left(R^2-2\rC^2+e^{-\tfrac{R^2}{\rC^2}}(R^2+2\rC^2)\right),\\
\eta_\text{\tiny (cuboid)}&=\frac{32\lcsl m^2 \rC^4}{L^4H^2m_0^2}\left(1-e^{-\tfrac{H^2}{4\rC^2}}\right)\cdot\\
&\cdot\left(1-e^{-\tfrac{L^2}{4\rC^2}}-\frac{L\sqrt{\pi}}{2r_C}\erf(\tfrac{L}{2\rC})\right)^2,\\
\eta_\text{\tiny (cylinder)}&=\frac{16m^2\rC^2\lcsl}{ H^2m_0^2L^2}{\left(1-e^{-\tfrac{H^2}{4\rC^2}}\right)}\cdot\\
&\cdot\left(1-e^{-\tfrac{L^2}{2\rC^2}}\left(\besseli_0\left(\tfrac{L^2}{2\rC^2}\right)+\besseli_1\left(\tfrac{L^2}{2\rC^2}\right)\right)\right),
\eqali
 {where $\besseli_i(x)$ denotes the modified Bessel function.}

In Fig.~\ref{thermalvsnonthermal} we compare these contributions. As one can see, for small values of the mass, corresponding to a system whose spatial dimension is smaller than $\rC$, the CSL diffusion rate depends on the shape in a negligible way. Conversely, for larger masses, or equivalently when the dimensions of the system exceed $\rC$, the shape of the system plays a role. The most favorable case is given by the cuboidal geometry, as it can be concluded from Fig.~\ref{cubvscyl}, where the cuboidal geometry is compared to the cylindrical one for different values of $L$  {(this means that the heights of the two systems will be different).} For $L\ll\rC$, there is no significant difference between the two geometries, as it is for the sphere. For $L\gg\rC$, the cuboidal geometry has a larger diffusion constant $\eta$, which leads to a stronger bound on $\lambda$.  {One can also consider an alternative analysis where the cuboidal and cylindrical geometries are compared for different values of $H$. In such a case no appreciable differences emerge between the two geometries. }

\section{Multilayer technique: a study case}
\label{App2}

{To better understand the enhancement that can be achieved with the multilayer technique, for the sake of simplicity let us compare the single layer case ($\Nlay=0$) with the two layer case ($\Nlay=1$) with $\muB=0$. The situation is represented in Fig.~\ref{plottotab}. The Fourier transform of the mass densities can be derived from Eq.~\eqref{eqdensity}, where in the single mass case, one has $H=a_1$ and $b_1=0$; while in the two-masses case $H=(2a_2+b_2)$ with $a_1=2a_2$, and $b_2\neq0$. Thus, one obtains
\bqali\label{compare12}
\tilde \mu_z^{(\Nlay=0)}(k_z)&=\frac{2\muA}{k_z}\sin(k_z a_2),\\
\tilde \mu_z^{(\Nlay=1)}(k_z)&=\frac{4\muA\sin(\tfrac12k_z a_2)}{k_z}\cos(\tfrac12k_z(a_2+b_2)),
\eqali
where, to make the comparison more direct, we express both the expressions in terms of $a_2$ and $b_2$. 
Due to the different geometry, in the second expression a cosine appears, and, by suitably choosing the values of $b_2$, this gives an enhancement of the CSL effect. 
In the limit of $\rC\to+\infty$, due to the presence of the Gaussian factor in Eq.~\eqref{defIz}, only small values of $k_z$ contribute to $\mathcal I_z$. This is the case where the collapse noise sees the system as point-like, regardless its geometry. In such a limit, the expressions in Eq.~\eqref{compare12} take the same value: $\lim_{k_z\to0}\tilde \mu_z^{(\Nlay=0,1)}(k_z)=2\muA a_2$, and thus the corresponding bounds are the same. Conversely, for $\rC\to0$, one needs to go back to the integrals in Eq.~\eqref{defIz}, which in our case can be computed exactly and read:
\bqali
\mathcal I_z^{(\Nlay=0)}&=\frac{2\sqrt{\pi}\mu_\text{\tiny A}^2}{\rC}\left(1-e^{-\tfrac{a_2^2}{\rC^2}}\right),\\
\mathcal I_z^{(\Nlay=1)}&=\frac{4\sqrt{\pi}\mu_\text{\tiny A}^2}{\rC}\left(1-e^{-\tfrac{a_2^2}{4\rC^2}}+\tfrac12 f_\text{\tiny geom}\right),
\eqali
where the first expression is in agreement with Eq.~\eqref{defIzN0} with $H=2a_2$, and where we defined
\bq
f_\text{\tiny geom}=2e^{-\tfrac{(a_2+b_2)^2}{4\rC^2}}-e^{-\tfrac{b_2^2}{4\rC^2}}-e^{-\tfrac{(2a_2+b_2)^2}{4\rC^2}},
\eq
which is a geometrical factor explicitly depending on $b_2$. For $b_2\neq0$, in the limit $\rC\to0$, 
one finds that the effect in the two layer case is twice that in the single layer case: $\lim_{\rC\to0}\mathcal I_z^{(\Nlay=1)}=2\times\lim_{\rC\to0}\mathcal I_z^{(\Nlay=0)}=4\sqrt{\pi}\muA^2/\rC$. This enhancement is due to a geometrical factor, which is different in the two configurations. Something similar happens when the cuboidal and the cylindrical geometries are compared for small values of $\rC$, as it is shown in Appendix \ref{App.levitated}. For $b_2=0$, one finds that $\lim_{\rC\to0}f_\text{\tiny geom}=-1$, and the single layer result is recovered as expected. One should note, however, that in the limit  $\rC\to0$ one goes beyond the limits of validity of the approximations used in the text. }


\begin{thebibliography}{44}%
\makeatletter
\providecommand \@ifxundefined [1]{%
 \@ifx{#1\undefined}
}%
\providecommand \@ifnum [1]{%
 \ifnum #1\expandafter \@firstoftwo
 \else \expandafter \@secondoftwo
 \fi
}%
\providecommand \@ifx [1]{%
 \ifx #1\expandafter \@firstoftwo
 \else \expandafter \@secondoftwo
 \fi
}%
\providecommand \natexlab [1]{#1}%
\providecommand \enquote  [1]{``#1''}%
\providecommand \bibnamefont  [1]{#1}%
\providecommand \bibfnamefont [1]{#1}%
\providecommand \citenamefont [1]{#1}%
\providecommand \href@noop [0]{\@secondoftwo}%
\providecommand \href [0]{\begingroup \@sanitize@url \@href}%
\providecommand \@href[1]{\@@startlink{#1}\@@href}%
\providecommand \@@href[1]{\endgroup#1\@@endlink}%
\providecommand \@sanitize@url [0]{\catcode `\\12\catcode `\$12\catcode
  `\&12\catcode `\#12\catcode `\^12\catcode `\_12\catcode `\%12\relax}%
\providecommand \@@startlink[1]{}%
\providecommand \@@endlink[0]{}%
\providecommand \url  [0]{\begingroup\@sanitize@url \@url }%
\providecommand \@url [1]{\endgroup\@href {#1}{\urlprefix }}%
\providecommand \urlprefix  [0]{URL }%
\providecommand \Eprint [0]{\href }%
\providecommand \doibase [0]{http://dx.doi.org/}%
\providecommand \selectlanguage [0]{\@gobble}%
\providecommand \bibinfo  [0]{\@secondoftwo}%
\providecommand \bibfield  [0]{\@secondoftwo}%
\providecommand \translation [1]{[#1]}%
\providecommand \BibitemOpen [0]{}%
\providecommand \bibitemStop [0]{}%
\providecommand \bibitemNoStop [0]{.\EOS\space}%
\providecommand \EOS [0]{\spacefactor3000\relax}%
\providecommand \BibitemShut  [1]{\csname bibitem#1\endcsname}%
\let\auto@bib@innerbib\@empty
\bibitem [{\citenamefont {Sinha}\ \emph {et~al.}(2010)\citenamefont {Sinha}
  \emph {et~al.}}]{Sinha:2010aa}%
  \BibitemOpen
  \bibfield  {author} {\bibinfo {author} {\bibfnamefont {U.}~\bibnamefont
  {Sinha}} \emph {et~al.},\ }\href
  {http://science.sciencemag.org/content/329/5990/418} {\bibfield  {journal}
  {\bibinfo  {journal} {Science}\ }\textbf {\bibinfo {volume} {329}},\ \bibinfo
  {pages} {418} (\bibinfo {year} {2010})}\BibitemShut {NoStop}%
\bibitem [{\citenamefont {Hensen}\ \emph {et~al.}(2015)\citenamefont {Hensen}
  \emph {et~al.}}]{Hensen:2015aa}%
  \BibitemOpen
  \bibfield  {author} {\bibinfo {author} {\bibfnamefont {B.}~\bibnamefont
  {Hensen}} \emph {et~al.},\ }\href {http://dx.doi.org/10.1038/nature15759}
  {\bibfield  {journal} {\bibinfo  {journal} {Nature}\ }\textbf {\bibinfo
  {volume} {526}},\ \bibinfo {pages} {682 EP } (\bibinfo {year}
  {2015})}\BibitemShut {NoStop}%
\bibitem [{\citenamefont {Procopio}\ \emph {et~al.}(2017)\citenamefont
  {Procopio} \emph {et~al.}}]{Procopio:2017aa}%
  \BibitemOpen
  \bibfield  {author} {\bibinfo {author} {\bibfnamefont {L.~M.}\ \bibnamefont
  {Procopio}} \emph {et~al.},\ }\href {http://dx.doi.org/10.1038/ncomms15044}
  {\bibfield  {journal} {\bibinfo  {journal} {Nat.~Commun.}\ }\textbf {\bibinfo
  {volume} {8}},\ \bibinfo {pages} {15044 EP } (\bibinfo {year}
  {2017})}\BibitemShut {NoStop}%
\bibitem [{\citenamefont {Kovachy}\ \emph {et~al.}(2015)\citenamefont {Kovachy}
  \emph {et~al.}}]{Kovachy:2015ab}%
  \BibitemOpen
  \bibfield  {author} {\bibinfo {author} {\bibfnamefont {T.}~\bibnamefont
  {Kovachy}} \emph {et~al.},\ }\href
  {https://link.aps.org/doi/10.1103/PhysRevLett.114.143004} {\bibfield
  {journal} {\bibinfo  {journal} {Phys.~Rev.~Lett.}\ }\textbf {\bibinfo
  {volume} {114}},\ \bibinfo {pages} {143004} (\bibinfo {year}
  {2015})}\BibitemShut {NoStop}%
\bibitem [{\citenamefont {Usenko}\ \emph {et~al.}(2011)\citenamefont {Usenko},
  \citenamefont {Vinante}, \citenamefont {Wijts},\ and\ \citenamefont
  {Oosterkamp}}]{Usenko:2011aa}%
  \BibitemOpen
  \bibfield  {author} {\bibinfo {author} {\bibfnamefont {O.}~\bibnamefont
  {Usenko}}, \bibinfo {author} {\bibfnamefont {A.}~\bibnamefont {Vinante}},
  \bibinfo {author} {\bibfnamefont {G.}~\bibnamefont {Wijts}}, \ and\ \bibinfo
  {author} {\bibfnamefont {T.~H.}\ \bibnamefont {Oosterkamp}},\ }\href
  {http://dx.doi.org/10.1063/1.3570628} {\bibfield  {journal} {\bibinfo
  {journal} {App.~Phys.~Lett.}\ }\textbf {\bibinfo {volume} {98}},\ \bibinfo
  {pages} {133105} (\bibinfo {year} {2011})}\BibitemShut {NoStop}%
\bibitem [{\citenamefont {Vinante}\ \emph {et~al.}(2006)\citenamefont {Vinante}
  \emph {et~al.}}]{Vinante:2006aa}%
  \BibitemOpen
  \bibfield  {author} {\bibinfo {author} {\bibfnamefont {A.}~\bibnamefont
  {Vinante}} \emph {et~al.} (\bibinfo {collaboration} {AURIGA Collaboration}),\
  }\href {http://stacks.iop.org/0264-9381/23/i=8/a=S14} {\bibfield  {journal}
  {\bibinfo  {journal} {Class.~Quantum Grav.}\ }\textbf {\bibinfo {volume}
  {23}},\ \bibinfo {pages} {S103} (\bibinfo {year} {2006})}\BibitemShut
  {NoStop}%
\bibitem [{\citenamefont {Abbott}\ \emph
  {et~al.}(2016{\natexlab{a}})\citenamefont {Abbott} \emph
  {et~al.}}]{Abbott:2016ab}%
  \BibitemOpen
  \bibfield  {author} {\bibinfo {author} {\bibfnamefont {B.~P.}\ \bibnamefont
  {Abbott}} \emph {et~al.} (\bibinfo {collaboration} {LIGO Scientific
  Collaboration and Virgo Collaboration}),\ }\href
  {http://link.aps.org/doi/10.1103/PhysRevLett.116.131103} {\bibfield
  {journal} {\bibinfo  {journal} {Phys.~Rev.~Lett.}\ }\textbf {\bibinfo
  {volume} {116}},\ \bibinfo {pages} {131103} (\bibinfo {year}
  {2016}{\natexlab{a}})}\BibitemShut {NoStop}%
\bibitem [{\citenamefont {Abbott}\ \emph
  {et~al.}(2016{\natexlab{b}})\citenamefont {Abbott} \emph
  {et~al.}}]{Abbott:2016aa}%
  \BibitemOpen
  \bibfield  {author} {\bibinfo {author} {\bibfnamefont {B.~P.}\ \bibnamefont
  {Abbott}} \emph {et~al.} (\bibinfo {collaboration} {LIGO Scientific
  Collaboration and Virgo Collaboration}),\ }\href
  {http://link.aps.org/doi/10.1103/PhysRevLett.116.061102} {\bibfield
  {journal} {\bibinfo  {journal} {Phys.~Rev.~Lett.}\ }\textbf {\bibinfo
  {volume} {116}},\ \bibinfo {pages} {061102} (\bibinfo {year}
  {2016}{\natexlab{b}})}\BibitemShut {NoStop}%
\bibitem [{\citenamefont {Armano}\ \emph {et~al.}(2016)\citenamefont {Armano}
  \emph {et~al.}}]{Armano:2016aa}%
  \BibitemOpen
  \bibfield  {author} {\bibinfo {author} {\bibfnamefont {M.}~\bibnamefont
  {Armano}} \emph {et~al.},\ }\href
  {http://link.aps.org/doi/10.1103/PhysRevLett.116.231101} {\bibfield
  {journal} {\bibinfo  {journal} {Phys.~Rev.~Lett.}\ }\textbf {\bibinfo
  {volume} {116}},\ \bibinfo {pages} {231101} (\bibinfo {year}
  {2016})}\BibitemShut {NoStop}%
\bibitem [{\citenamefont {Armano}\ \emph {et~al.}(2018)\citenamefont {Armano}
  \emph {et~al.}}]{Armano:2018aa}%
  \BibitemOpen
  \bibfield  {author} {\bibinfo {author} {\bibfnamefont {M.}~\bibnamefont
  {Armano}} \emph {et~al.},\ }\href
  {https://link.aps.org/doi/10.1103/PhysRevLett.120.061101} {\bibfield
  {journal} {\bibinfo  {journal} {Phys.~Rev.~Lett.}\ }\textbf {\bibinfo
  {volume} {120}},\ \bibinfo {pages} {061101} (\bibinfo {year}
  {2018})}\BibitemShut {NoStop}%
\bibitem [{\citenamefont {Aalseth}\ \emph {et~al.}(1999)\citenamefont {Aalseth}
  \emph {et~al.}}]{Aalseth:1999aa}%
  \BibitemOpen
  \bibfield  {author} {\bibinfo {author} {\bibfnamefont {C.~E.}\ \bibnamefont
  {Aalseth}} \emph {et~al.} (\bibinfo {collaboration} {The IGEX
  Collaboration}),\ }\href {https://link.aps.org/doi/10.1103/PhysRevC.59.2108}
  {\bibfield  {journal} {\bibinfo  {journal} {Phys.~Rev.~C}\ }\textbf {\bibinfo
  {volume} {59}},\ \bibinfo {pages} {2108} (\bibinfo {year}
  {1999})}\BibitemShut {NoStop}%
\bibitem [{\citenamefont {Adler}\ and\ \citenamefont
  {Vinante}(2018)}]{Adler:2018aa}%
  \BibitemOpen
  \bibfield  {author} {\bibinfo {author} {\bibfnamefont {S.~L.}\ \bibnamefont
  {Adler}}\ and\ \bibinfo {author} {\bibfnamefont {A.}~\bibnamefont
  {Vinante}},\ }\href {https://link.aps.org/doi/10.1103/PhysRevA.97.052119}
  {\bibfield  {journal} {\bibinfo  {journal} {Phys.~Rev.~A}\ }\textbf {\bibinfo
  {volume} {97}},\ \bibinfo {pages} {052119} (\bibinfo {year}
  {2018})}\BibitemShut {NoStop}%
\bibitem [{\citenamefont {Bahrami}(2018)}]{Bahrami:2018aa}%
  \BibitemOpen
  \bibfield  {author} {\bibinfo {author} {\bibfnamefont {M.}~\bibnamefont
  {Bahrami}},\ }\href {https://link.aps.org/doi/10.1103/PhysRevA.97.052118}
  {\bibfield  {journal} {\bibinfo  {journal} {Phys.~Rev.~A}\ }\textbf {\bibinfo
  {volume} {97}},\ \bibinfo {pages} {052118} (\bibinfo {year}
  {2018})}\BibitemShut {NoStop}%
\bibitem [{\citenamefont {Vinante}\ \emph {et~al.}(2016)\citenamefont {Vinante}
  \emph {et~al.}}]{Vinante:2016aa}%
  \BibitemOpen
  \bibfield  {author} {\bibinfo {author} {\bibfnamefont {A.}~\bibnamefont
  {Vinante}} \emph {et~al.},\ }\href
  {http://link.aps.org/doi/10.1103/PhysRevLett.116.090402} {\bibfield
  {journal} {\bibinfo  {journal} {Phys.~Rev.~Lett.}\ }\textbf {\bibinfo
  {volume} {116}},\ \bibinfo {pages} {090402} (\bibinfo {year}
  {2016})}\BibitemShut {NoStop}%
\bibitem [{\citenamefont {Vinante}\ \emph {et~al.}(2017)\citenamefont
  {Vinante}, \citenamefont {Mezzena}, \citenamefont {Falferi}, \citenamefont
  {Carlesso},\ and\ \citenamefont {Bassi}}]{Vinante:2017aa}%
  \BibitemOpen
  \bibfield  {author} {\bibinfo {author} {\bibfnamefont {A.}~\bibnamefont
  {Vinante}}, \bibinfo {author} {\bibfnamefont {R.}~\bibnamefont {Mezzena}},
  \bibinfo {author} {\bibfnamefont {P.}~\bibnamefont {Falferi}}, \bibinfo
  {author} {\bibfnamefont {M.}~\bibnamefont {Carlesso}}, \ and\ \bibinfo
  {author} {\bibfnamefont {A.}~\bibnamefont {Bassi}},\ }\href
  {https://link.aps.org/doi/10.1103/PhysRevLett.119.110401} {\bibfield
  {journal} {\bibinfo  {journal} {Phys.~Rev.~Lett.}\ }\textbf {\bibinfo
  {volume} {119}},\ \bibinfo {pages} {110401} (\bibinfo {year}
  {2017})}\BibitemShut {NoStop}%
\bibitem [{\citenamefont {Carlesso}\ \emph {et~al.}(2016)\citenamefont
  {Carlesso}, \citenamefont {Bassi}, \citenamefont {Falferi},\ and\
  \citenamefont {Vinante}}]{Carlesso:2016aa}%
  \BibitemOpen
  \bibfield  {author} {\bibinfo {author} {\bibfnamefont {M.}~\bibnamefont
  {Carlesso}}, \bibinfo {author} {\bibfnamefont {A.}~\bibnamefont {Bassi}},
  \bibinfo {author} {\bibfnamefont {P.}~\bibnamefont {Falferi}}, \ and\
  \bibinfo {author} {\bibfnamefont {A.}~\bibnamefont {Vinante}},\ }\href
  {http://link.aps.org/doi/10.1103/PhysRevD.94.124036} {\bibfield  {journal}
  {\bibinfo  {journal} {Phys.~Rev.~D}\ }\textbf {\bibinfo {volume} {94}},\
  \bibinfo {pages} {124036} (\bibinfo {year} {2016})}\BibitemShut {NoStop}%
\bibitem [{\citenamefont {Helou}\ \emph {et~al.}(2017)\citenamefont {Helou},
  \citenamefont {Slagmolen}, \citenamefont {McClelland},\ and\ \citenamefont
  {Chen}}]{Helou:2017aa}%
  \BibitemOpen
  \bibfield  {author} {\bibinfo {author} {\bibfnamefont {B.}~\bibnamefont
  {Helou}}, \bibinfo {author} {\bibfnamefont {B.~J.~J.}\ \bibnamefont
  {Slagmolen}}, \bibinfo {author} {\bibfnamefont {D.~E.}\ \bibnamefont
  {McClelland}}, \ and\ \bibinfo {author} {\bibfnamefont {Y.}~\bibnamefont
  {Chen}},\ }\href {https://link.aps.org/doi/10.1103/PhysRevD.95.084054}
  {\bibfield  {journal} {\bibinfo  {journal} {Phys.~Rev.~D}\ }\textbf {\bibinfo
  {volume} {95}},\ \bibinfo {pages} {084054} (\bibinfo {year}
  {2017})}\BibitemShut {NoStop}%
\bibitem [{\citenamefont {Piscicchia}\ \emph {et~al.}(2017)\citenamefont
  {Piscicchia} \emph {et~al.}}]{Piscicchia:2017aa}%
  \BibitemOpen
  \bibfield  {author} {\bibinfo {author} {\bibfnamefont {K.}~\bibnamefont
  {Piscicchia}} \emph {et~al.},\ }\href
  {http://www.mdpi.com/1099-4300/19/7/319} {\bibfield  {journal} {\bibinfo
  {journal} {Entropy}\ }\textbf {\bibinfo {volume} {19}} (\bibinfo {year}
  {2017})}\BibitemShut {NoStop}%
\bibitem [{\citenamefont {Bassi}\ and\ \citenamefont
  {Ghirardi}(2003)}]{Bassi:2003aa}%
  \BibitemOpen
  \bibfield  {author} {\bibinfo {author} {\bibfnamefont {A.}~\bibnamefont
  {Bassi}}\ and\ \bibinfo {author} {\bibfnamefont {G.~C.}\ \bibnamefont
  {Ghirardi}},\ }\href {\doibase
  http://dx.doi.org/10.1016/S0370-1573(03)00103-0} {\bibfield  {journal}
  {\bibinfo  {journal} {Phys.~Rep.}\ }\textbf {\bibinfo {volume} {379}},\
  \bibinfo {pages} {257 } (\bibinfo {year} {2003})}\BibitemShut {NoStop}%
\bibitem [{\citenamefont {Bassi}\ \emph {et~al.}(2013)\citenamefont {Bassi}
  \emph {et~al.}}]{Bassi:2013aa}%
  \BibitemOpen
  \bibfield  {author} {\bibinfo {author} {\bibfnamefont {A.}~\bibnamefont
  {Bassi}} \emph {et~al.},\ }\href
  {http://link.aps.org/doi/10.1103/RevModPhys.85.471} {\bibfield  {journal}
  {\bibinfo  {journal} {Rev.~Mod.~Phys.}\ }\textbf {\bibinfo {volume} {85}},\
  \bibinfo {pages} {471} (\bibinfo {year} {2013})}\BibitemShut {NoStop}%
\bibitem [{\citenamefont {Ghirardi}\ \emph {et~al.}(1986)\citenamefont
  {Ghirardi}, \citenamefont {Rimini},\ and\ \citenamefont
  {Weber}}]{Ghirardi:1986aa}%
  \BibitemOpen
  \bibfield  {author} {\bibinfo {author} {\bibfnamefont {G.~C.}\ \bibnamefont
  {Ghirardi}}, \bibinfo {author} {\bibfnamefont {A.}~\bibnamefont {Rimini}}, \
  and\ \bibinfo {author} {\bibfnamefont {T.}~\bibnamefont {Weber}},\ }\href
  {http://link.aps.org/doi/10.1103/PhysRevD.34.470} {\bibfield  {journal}
  {\bibinfo  {journal} {Phys.~Rev.~D}\ }\textbf {\bibinfo {volume} {34}},\
  \bibinfo {pages} {470} (\bibinfo {year} {1986})}\BibitemShut {NoStop}%
\bibitem [{\citenamefont {Adler}(2007{\natexlab{a}})}]{Adler:2007ab}%
  \BibitemOpen
  \bibfield  {author} {\bibinfo {author} {\bibfnamefont {S.~L.}\ \bibnamefont
  {Adler}},\ }\href {http://stacks.iop.org/1751-8121/40/i=12/a=S03} {\bibfield
  {journal} {\bibinfo  {journal} {J. Phys. A}\ }\textbf {\bibinfo {volume}
  {40}},\ \bibinfo {pages} {2935} (\bibinfo {year}
  {2007}{\natexlab{a}})}\BibitemShut {NoStop}%
\bibitem [{\citenamefont {Adler}(2007{\natexlab{b}})}]{Adler:2007ac}%
  \BibitemOpen
  \bibfield  {author} {\bibinfo {author} {\bibfnamefont {S.~L.}\ \bibnamefont
  {Adler}},\ }\href {http://stacks.iop.org/1751-8121/40/i=44/a=C01} {\bibfield
  {journal} {\bibinfo  {journal} {J. Phys. A}\ }\textbf {\bibinfo {volume}
  {40}},\ \bibinfo {pages} {13501} (\bibinfo {year}
  {2007}{\natexlab{b}})}\BibitemShut {NoStop}%
\bibitem [{\citenamefont {Eibenberger}\ \emph {et~al.}(2013)\citenamefont
  {Eibenberger} \emph {et~al.}}]{Eibenberger:2013aa}%
  \BibitemOpen
  \bibfield  {author} {\bibinfo {author} {\bibfnamefont {S.}~\bibnamefont
  {Eibenberger}} \emph {et~al.},\ }\href {http://dx.doi.org/10.1039/C3CP51500A}
  {\bibfield  {journal} {\bibinfo  {journal} {Phys.~Chem.~Chem.~Phys.}\
  }\textbf {\bibinfo {volume} {15}},\ \bibinfo {pages} {14696} (\bibinfo {year}
  {2013})}\BibitemShut {NoStop}%
\bibitem [{\citenamefont {Hornberger}\ \emph {et~al.}(2004)\citenamefont
  {Hornberger}, \citenamefont {Sipe},\ and\ \citenamefont
  {Arndt}}]{Hornberger:2004aa}%
  \BibitemOpen
  \bibfield  {author} {\bibinfo {author} {\bibfnamefont {K.}~\bibnamefont
  {Hornberger}}, \bibinfo {author} {\bibfnamefont {J.~E.}\ \bibnamefont
  {Sipe}}, \ and\ \bibinfo {author} {\bibfnamefont {M.}~\bibnamefont {Arndt}},\
  }\href {\doibase 10.1103/PhysRevA.70.053608} {\bibfield  {journal} {\bibinfo
  {journal} {Phys.~Rev.~A}\ }\textbf {\bibinfo {volume} {70}},\ \bibinfo
  {pages} {053608} (\bibinfo {year} {2004})}\BibitemShut {NoStop}%
\bibitem [{\citenamefont {Toro{\v s}}\ \emph {et~al.}(2017)\citenamefont
  {Toro{\v s}}, \citenamefont {Gasbarri},\ and\ \citenamefont
  {Bassi}}]{Toros:2017aa}%
  \BibitemOpen
  \bibfield  {author} {\bibinfo {author} {\bibfnamefont {M.}~\bibnamefont
  {Toro{\v s}}}, \bibinfo {author} {\bibfnamefont {G.}~\bibnamefont
  {Gasbarri}}, \ and\ \bibinfo {author} {\bibfnamefont {A.}~\bibnamefont
  {Bassi}},\ }\href
  {http://www.sciencedirect.com/science/article/pii/S0375960117309465}
  {\bibfield  {journal} {\bibinfo  {journal} {Phys.~Lett.~A}\ }\textbf
  {\bibinfo {volume} {381}},\ \bibinfo {pages} {3921 } (\bibinfo {year}
  {2017})}\BibitemShut {NoStop}%
\bibitem [{\citenamefont {Toro{\v s}}\ and\ \citenamefont
  {Bassi}(2018)}]{Toros:2018aa}%
  \BibitemOpen
  \bibfield  {author} {\bibinfo {author} {\bibfnamefont {M.}~\bibnamefont
  {Toro{\v s}}}\ and\ \bibinfo {author} {\bibfnamefont {A.}~\bibnamefont
  {Bassi}},\ }\href {http://stacks.iop.org/1751-8121/51/i=11/a=115302}
  {\bibfield  {journal} {\bibinfo  {journal} {J.~Phys.~A}\ }\textbf {\bibinfo
  {volume} {51}},\ \bibinfo {pages} {115302} (\bibinfo {year}
  {2018})}\BibitemShut {NoStop}%
\bibitem [{\citenamefont {Lee}\ \emph {et~al.}(2011)\citenamefont {Lee} \emph
  {et~al.}}]{Lee:2011aa}%
  \BibitemOpen
  \bibfield  {author} {\bibinfo {author} {\bibfnamefont {K.~C.}\ \bibnamefont
  {Lee}} \emph {et~al.},\ }\href
  {http://science.sciencemag.org/content/334/6060/1253} {\bibfield  {journal}
  {\bibinfo  {journal} {Science}\ }\textbf {\bibinfo {volume} {334}},\ \bibinfo
  {pages} {1253} (\bibinfo {year} {2011})}\BibitemShut {NoStop}%
\bibitem [{\citenamefont {Belli}\ \emph {et~al.}(2016)\citenamefont {Belli}
  \emph {et~al.}}]{Belli:2016aa}%
  \BibitemOpen
  \bibfield  {author} {\bibinfo {author} {\bibfnamefont {S.}~\bibnamefont
  {Belli}} \emph {et~al.},\ }\href
  {http://link.aps.org/doi/10.1103/PhysRevA.94.012108} {\bibfield  {journal}
  {\bibinfo  {journal} {Phys. Rev. A}\ }\textbf {\bibinfo {volume} {94}},\
  \bibinfo {pages} {012108} (\bibinfo {year} {2016})}\BibitemShut {NoStop}%
\bibitem [{\citenamefont {Bahrami}\ \emph {et~al.}(2014)\citenamefont {Bahrami}
  \emph {et~al.}}]{Bahrami:2014aa}%
  \BibitemOpen
  \bibfield  {author} {\bibinfo {author} {\bibfnamefont {M.}~\bibnamefont
  {Bahrami}} \emph {et~al.},\ }\href
  {http://link.aps.org/doi/10.1103/PhysRevLett.112.210404} {\bibfield
  {journal} {\bibinfo  {journal} {Phys.~Rev.~Lett.}\ }\textbf {\bibinfo
  {volume} {112}},\ \bibinfo {pages} {210404} (\bibinfo {year}
  {2014})}\BibitemShut {NoStop}%
\bibitem [{\citenamefont {Nimmrichter}\ \emph {et~al.}(2014)\citenamefont
  {Nimmrichter}, \citenamefont {Hornberger},\ and\ \citenamefont
  {Hammerer}}]{Nimmrichter:2014aa}%
  \BibitemOpen
  \bibfield  {author} {\bibinfo {author} {\bibfnamefont {S.}~\bibnamefont
  {Nimmrichter}}, \bibinfo {author} {\bibfnamefont {K.}~\bibnamefont
  {Hornberger}}, \ and\ \bibinfo {author} {\bibfnamefont {K.}~\bibnamefont
  {Hammerer}},\ }\href {http://link.aps.org/doi/10.1103/PhysRevLett.113.020405}
  {\bibfield  {journal} {\bibinfo  {journal} {Phys.~Rev.~Lett.}\ }\textbf
  {\bibinfo {volume} {113}},\ \bibinfo {pages} {020405} (\bibinfo {year}
  {2014})}\BibitemShut {NoStop}%
\bibitem [{\citenamefont {Di\'osi}(2015)}]{Diosi:2015ab}%
  \BibitemOpen
  \bibfield  {author} {\bibinfo {author} {\bibfnamefont {L.}~\bibnamefont
  {Di\'osi}},\ }\href {http://link.aps.org/doi/10.1103/PhysRevLett.114.050403}
  {\bibfield  {journal} {\bibinfo  {journal} {Phys.~Rev.~Lett.}\ }\textbf
  {\bibinfo {volume} {114}},\ \bibinfo {pages} {050403} (\bibinfo {year}
  {2015})}\BibitemShut {NoStop}%
\bibitem [{\citenamefont {Bilardello}\ \emph {et~al.}(2016)\citenamefont
  {Bilardello}, \citenamefont {Donadi}, \citenamefont {Vinante},\ and\
  \citenamefont {Bassi}}]{Bilardello:2016aa}%
  \BibitemOpen
  \bibfield  {author} {\bibinfo {author} {\bibfnamefont {M.}~\bibnamefont
  {Bilardello}}, \bibinfo {author} {\bibfnamefont {S.}~\bibnamefont {Donadi}},
  \bibinfo {author} {\bibfnamefont {A.}~\bibnamefont {Vinante}}, \ and\
  \bibinfo {author} {\bibfnamefont {A.}~\bibnamefont {Bassi}},\ }\href
  {http://www.sciencedirect.com/science/article/pii/S0378437116304095}
  {\bibfield  {journal} {\bibinfo  {journal} {Physica A}\ }\textbf {\bibinfo
  {volume} {462}},\ \bibinfo {pages} {764 } (\bibinfo {year}
  {2016})}\BibitemShut {NoStop}%
\bibitem [{\citenamefont {Bassi}\ \emph {et~al.}(2017)\citenamefont {Bassi},
  \citenamefont {Gro{\ss}ardt},\ and\ \citenamefont {Ulbricht}}]{Bassi:2017aa}%
  \BibitemOpen
  \bibfield  {author} {\bibinfo {author} {\bibfnamefont {A.}~\bibnamefont
  {Bassi}}, \bibinfo {author} {\bibfnamefont {A.}~\bibnamefont {Gro{\ss}ardt}},
  \ and\ \bibinfo {author} {\bibfnamefont {H.}~\bibnamefont {Ulbricht}},\
  }\href {http://stacks.iop.org/0264-9381/34/i=19/a=193002} {\bibfield
  {journal} {\bibinfo  {journal} {Class.~Quantum Grav.}\ }\textbf {\bibinfo
  {volume} {34}},\ \bibinfo {pages} {193002} (\bibinfo {year}
  {2017})}\BibitemShut {NoStop}%
\bibitem [{\citenamefont {Goldwater}\ \emph {et~al.}(2016)\citenamefont
  {Goldwater}, \citenamefont {Paternostro},\ and\ \citenamefont
  {Barker}}]{Goldwater:2016aa}%
  \BibitemOpen
  \bibfield  {author} {\bibinfo {author} {\bibfnamefont {D.}~\bibnamefont
  {Goldwater}}, \bibinfo {author} {\bibfnamefont {M.}~\bibnamefont
  {Paternostro}}, \ and\ \bibinfo {author} {\bibfnamefont {P.~F.}\ \bibnamefont
  {Barker}},\ }\href {\doibase 10.1103/PhysRevA.94.010104} {\bibfield
  {journal} {\bibinfo  {journal} {Phys. Rev. A}\ }\textbf {\bibinfo {volume}
  {94}},\ \bibinfo {pages} {010104} (\bibinfo {year} {2016})}\BibitemShut
  {NoStop}%
\bibitem [{\citenamefont {McMillen}\ \emph {et~al.}(2017)\citenamefont
  {McMillen} \emph {et~al.}}]{McMillen:2017aa}%
  \BibitemOpen
  \bibfield  {author} {\bibinfo {author} {\bibfnamefont {S.}~\bibnamefont
  {McMillen}} \emph {et~al.},\ }\href
  {https://link.aps.org/doi/10.1103/PhysRevA.95.012132} {\bibfield  {journal}
  {\bibinfo  {journal} {Phys.~Rev.~A}\ }\textbf {\bibinfo {volume} {95}},\
  \bibinfo {pages} {012132} (\bibinfo {year} {2017})}\BibitemShut {NoStop}%
\bibitem [{\citenamefont {Schrinski}\ \emph {et~al.}(2017)\citenamefont
  {Schrinski}, \citenamefont {Stickler},\ and\ \citenamefont
  {Hornberger}}]{Schrinski:2017aa}%
  \BibitemOpen
  \bibfield  {author} {\bibinfo {author} {\bibfnamefont {B.}~\bibnamefont
  {Schrinski}}, \bibinfo {author} {\bibfnamefont {B.~A.}\ \bibnamefont
  {Stickler}}, \ and\ \bibinfo {author} {\bibfnamefont {K.}~\bibnamefont
  {Hornberger}},\ }\href {http://josab.osa.org/abstract.cfm?URI=josab-34-6-C1}
  {\bibfield  {journal} {\bibinfo  {journal} {J.~Opt.~Soc.~Am.~B}\ }\textbf
  {\bibinfo {volume} {34}},\ \bibinfo {pages} {C1} (\bibinfo {year}
  {2017})}\BibitemShut {NoStop}%
\bibitem [{\citenamefont {Carlesso}\ \emph
  {et~al.}(2018{\natexlab{a}})\citenamefont {Carlesso}, \citenamefont
  {Paternostro}, \citenamefont {Ulbricht}, \citenamefont {Vinante},\ and\
  \citenamefont {Bassi}}]{carlesso:2018ab}%
  \BibitemOpen
  \bibfield  {author} {\bibinfo {author} {\bibfnamefont {M.}~\bibnamefont
  {Carlesso}}, \bibinfo {author} {\bibfnamefont {M.}~\bibnamefont
  {Paternostro}}, \bibinfo {author} {\bibfnamefont {H.}~\bibnamefont
  {Ulbricht}}, \bibinfo {author} {\bibfnamefont {A.}~\bibnamefont {Vinante}}, \
  and\ \bibinfo {author} {\bibfnamefont {A.}~\bibnamefont {Bassi}},\ }\href
  {http://stacks.iop.org/1367-2630/20/i=8/a=083022} {\bibfield  {journal}
  {\bibinfo  {journal} {New J.~Phys.}\ }\textbf {\bibinfo {volume} {20}},\
  \bibinfo {pages} {083022} (\bibinfo {year} {2018}{\natexlab{a}})}\BibitemShut
  {NoStop}%
\bibitem [{\citenamefont {Smith}\ and\ \citenamefont
  {Lewin}(1984)}]{Smith:1984aa}%
  \BibitemOpen
  \bibfield  {author} {\bibinfo {author} {\bibfnamefont {P.~F.}\ \bibnamefont
  {Smith}}\ and\ \bibinfo {author} {\bibfnamefont {J.~D.}\ \bibnamefont
  {Lewin}},\ }\href@noop {} {\bibfield  {journal} {\bibinfo  {journal} {Acta
  Phys.~Polonica}\ }\textbf {\bibinfo {volume} {15}},\ \bibinfo {pages} {1201}
  (\bibinfo {year} {1984})}\BibitemShut {NoStop}%
\bibitem [{\citenamefont {Curceanu}\ \emph {et~al.}(2015)\citenamefont
  {Curceanu}, \citenamefont {Hiesmayr},\ and\ \citenamefont
  {Piscicchia}}]{Curceanu:2015aa}%
  \BibitemOpen
  \bibfield  {author} {\bibinfo {author} {\bibfnamefont {C.}~\bibnamefont
  {Curceanu}}, \bibinfo {author} {\bibfnamefont {B.~C.}\ \bibnamefont
  {Hiesmayr}}, \ and\ \bibinfo {author} {\bibfnamefont {K.}~\bibnamefont
  {Piscicchia}},\ }\href
  {http://www.ingentaconnect.com/content/asp/jap/2015/00000004/00000003/art00017}
  {\bibfield  {journal} {\bibinfo  {journal} {J.~Adv.~Phys.}\ }\textbf
  {\bibinfo {volume} {4}},\ \bibinfo {pages} {263} (\bibinfo {year}
  {2015})}\BibitemShut {NoStop}%
\bibitem [{\citenamefont {Lalo\"e}\ \emph {et~al.}(2014)\citenamefont
  {Lalo\"e}, \citenamefont {Mullin},\ and\ \citenamefont
  {Pearle}}]{Laloe:2014aa}%
  \BibitemOpen
  \bibfield  {author} {\bibinfo {author} {\bibfnamefont {F.}~\bibnamefont
  {Lalo\"e}}, \bibinfo {author} {\bibfnamefont {W.~J.}\ \bibnamefont {Mullin}},
  \ and\ \bibinfo {author} {\bibfnamefont {P.}~\bibnamefont {Pearle}},\ }\href
  {http://link.aps.org/doi/10.1103/PhysRevA.90.052119} {\bibfield  {journal}
  {\bibinfo  {journal} {Phys.~Rev.~A}\ }\textbf {\bibinfo {volume} {90}},\
  \bibinfo {pages} {052119} (\bibinfo {year} {2014})}\BibitemShut {NoStop}%
\bibitem [{\citenamefont {Nobakht}\ \emph {et~al.}(2018)\citenamefont
  {Nobakht}, \citenamefont {Carlesso}, \citenamefont {Donadi}, \citenamefont
  {Paternostro},\ and\ \citenamefont {Bassi}}]{Nobakht:2018aa}%
  \BibitemOpen
  \bibfield  {author} {\bibinfo {author} {\bibfnamefont {J.}~\bibnamefont
  {Nobakht}}, \bibinfo {author} {\bibfnamefont {M.}~\bibnamefont {Carlesso}},
  \bibinfo {author} {\bibfnamefont {S.}~\bibnamefont {Donadi}}, \bibinfo
  {author} {\bibfnamefont {M.}~\bibnamefont {Paternostro}}, \ and\ \bibinfo
  {author} {\bibfnamefont {A.}~\bibnamefont {Bassi}},\ }\href
  {https://arxiv.org/abs/1808.01143} {\bibfield  {journal} {\bibinfo  {journal}
  {ArXiv}\ } (\bibinfo {year} {2018})},\ \Eprint
  {http://arxiv.org/abs/1808.01143} {1808.01143} \BibitemShut {NoStop}%
\bibitem [{\citenamefont {Carlesso}\ \emph
  {et~al.}(2018{\natexlab{b}})\citenamefont {Carlesso}, \citenamefont
  {Ferialdi},\ and\ \citenamefont {Bassi}}]{Carlesso:2018aa}%
  \BibitemOpen
  \bibfield  {author} {\bibinfo {author} {\bibfnamefont {M.}~\bibnamefont
  {Carlesso}}, \bibinfo {author} {\bibfnamefont {L.}~\bibnamefont {Ferialdi}},
  \ and\ \bibinfo {author} {\bibfnamefont {A.}~\bibnamefont {Bassi}},\ }\href
  {https://arxiv.org/abs/1805.10100} {\bibfield  {journal} {\bibinfo  {journal}
  {ArXiv}\ } (\bibinfo {year} {2018}{\natexlab{b}})},\ \Eprint
  {http://arxiv.org/abs/1805.10100. (Accepted in European Physical Journal D)}
  {1805.10100. (Accepted in European Physical Journal D)\!\!} \BibitemShut
  {NoStop}%
\bibitem [{\citenamefont {Bassi}\ \emph {et~al.}(2010)\citenamefont {Bassi},
  \citenamefont {Deckert},\ and\ \citenamefont {Ferialdi}}]{Bassi:2010aa}%
  \BibitemOpen
  \bibfield  {author} {\bibinfo {author} {\bibfnamefont {A.}~\bibnamefont
  {Bassi}}, \bibinfo {author} {\bibfnamefont {D.-A.}\ \bibnamefont {Deckert}},
  \ and\ \bibinfo {author} {\bibfnamefont {L.}~\bibnamefont {Ferialdi}},\
  }\href {http://stacks.iop.org/0295-5075/92/i=5/a=50006} {\bibfield  {journal}
  {\bibinfo  {journal} {EPL (Europhysics Letters)}\ }\textbf {\bibinfo {volume}
  {92}},\ \bibinfo {pages} {50006} (\bibinfo {year} {2010})}\BibitemShut
  {NoStop}%
\end{thebibliography}
\end{document}